\documentclass[reprint,floatfix,amsmath,amssymb,aps]{revtex4-2}

\usepackage[utf8]{inputenc}
\usepackage{graphicx}
\usepackage{dcolumn}
\usepackage{bm}
\usepackage{textcomp}
\usepackage{placeins}
\usepackage{siunitx}
\usepackage[percent]{overpic}
\usepackage{subcaption}

\DeclareCaptionJustification{justified}{\justifying}
\usepackage{tikz}
\usepackage{ragged2e}

\newcommand{\RefHeading}{%
  \begin{center}{\bfseries\large REFERENCES}\end{center}\vspace{0.75em}}

\usepackage{placeins}     
\usepackage{dblfloatfix}  

\makeatletter
\def\fps@table{tbp}
\def\fps@figure{tbp}
\makeatother

\makeatletter
\newcommand{\BeginSupplement}{%
  \clearpage
  \twocolumngrid
  \def\ps@suppl{%
    \let\@oddfoot\@empty
    \let\@evenfoot\@empty
    \def\@oddhead{\hfill\thepage}%
    \def\@evenhead{\hfill\thepage}%
  }%
  \pagestyle{suppl}\thispagestyle{suppl}\markboth{}{}%
  \section*{Supplemental Material}%
  \setcounter{figure}{0}\renewcommand{\thefigure}{S\arabic{figure}}%
  \setcounter{table}{0}\renewcommand{\thetable}{S\arabic{table}}%
  \setcounter{equation}{0}\renewcommand{\theequation}{S\arabic{equation}}%
  \FloatBarrier
}
\makeatother

\begin{document}

\title[Novel Transformations of PbTiO$_3$ with Pressure and Temperature]{Novel Transformations of PbTiO$_3$ with Pressure and Temperature}

\makeatletter
\def\andname{}
\makeatother

\author{Husam Farraj$^{1,2}$, Stefano Racioppi$^{3,4}$, Gaston Garbarino$^{5}$, Muhtar Ahart$^{6}$,
Anshuman Mondal$^{5,7}$, Samuel G. Parra$^{5,8}$, Jesse S. Smith$^{9}$, R.~E. Cohen$^{10}$,
Eva Zurek$^{3}$, Jordi Cabana$^{1,11}$, Russell J. Hemley$^{1,6,11,12}$}

\affiliation{$^{1}$Department of Chemistry, University of Illinois Chicago, Chicago, IL 60607, USA}
\affiliation{$^{2}$Deutsches Elektronen-Synchrotron (DESY), 22607 Hamburg, Germany}
\affiliation{$^{3}$Department of Chemistry, State University of New York Buffalo, Buffalo, NY 14260, USA}
\affiliation{$^{4}$Department of Materials Science and Metallurgy, University of Cambridge, Cambridge CB3 0FS, UK}
\affiliation{$^{5}$European Synchrotron Radiation Facility, 38043 Grenoble Cedex 9, France}
\affiliation{$^{6}$Department of Physics, University of Illinois Chicago, Chicago, IL 60607, USA}
\affiliation{$^{7}$Department of Physics, National Institute of Technology, Agartala, India 799046}
\affiliation{$^{8}$Instituto de Diseño para la Fabricación y Producción Automatizada, MALTA Consolider Team,
Universitat Politècnica de València, 46022 València, Spain}
\affiliation{$^{9}$HPCAT, X-ray Science Division, Argonne National Laboratory, Lemont, IL 60439, USA}
\affiliation{$^{10}$Earth and Planets Laboratory, Carnegie Science, Washington, DC 20015, USA}
\affiliation{$^{11}$Materials Science Division, Argonne National Laboratory, Lemont, IL 60439, USA}
\affiliation{$^{12}$Department of Earth and Environmental Sciences, University of Illinois Chicago, Chicago, IL 60607, USA}

\date{\today}

\begin{abstract}
We investigated the behavior of lead titanate (PbTiO$_3$) up to 100~GPa, both at room temperature and upon laser heating, using synchrotron X-ray diffraction combined with density functional theory (DFT) computations. At the high pressure–temperature ($P$–$T$) conditions produced in laser-heated diamond-anvil cells, PbTiO$_3$ dissociates into PbO and TiO$_2$, consistent with our DFT computations showing that decomposition becomes enthalpically favored above 65~GPa. In contrast, on room temperature compression, PbTiO$_3$ persists in the tetragonal $I4/mcm$ phase up to at least 100~GPa. Laser heating produces distinct PbO phases: a compressed form of $\alpha$-PbO and a previously unreported $\delta$-PbO polymorph, both of which transform to $\beta$-PbO on decompression. The calculations predict that $\alpha$-PbO undergoes pressure-induced band gap closure, metallizing above 70 GPa, whereas the $\delta$ and $\beta$ phases remain semiconducting with a band gap above 1~eV even at megabar pressures. The experimental and confirming theoretical results reveal an unanticipated dimension of the behavior of PbTiO$_3$, showing that distinct equilibrium and metastable phases can be stabilized along different $P$–$T$ synthesis paths.
\end{abstract}

\maketitle
Perovskite oxides (ABO$_3$) exhibit diverse functionalities ranging from superconductivity to ferroelectricity~\cite{Zhang2023advances}. Lead titanate (PbTiO$_3$), a classic ferroelectric, is widely studied for prototypical device applications~\cite{Baudry2017,Shin2018,Moret2002,Yoon1993,Kim1995,Zhang2013,Kanda2012,Okuyama1991}. Early work by Shirane \textit{et al.} established its ferroelectric phase transition near 760~K using X-ray and neutron diffraction~\cite{Shirane1955}, while Burns \textit{et al.} demonstrated soft-mode behavior using Raman spectroscopy~\cite{Burns1970PRL}. These lattice instabilities result in a spontaneous polarization that stabilizes the tetragonal $P4mm$ phase with a high Curie temperature of 763~K~\cite{Shirane1955,Yoshiasa2016,Peng1991}, above which the cubic $Pm\bar{3}m$ phase remains stable up to the melting temperature near 1558~K~\cite{Grabmaier1976}. The seminal work by Cohen identified the role of Pb–O hybridization in stabilizing ferroelectricity~\cite{Cohen1990,Cohen1992}.

Subsequent high-pressure Raman studies by Sanjurjo \textit{et al.} revealed strong pressure-induced phonon softening~\cite{Sanjurjo1983}, consistent with a competition between long-range Coulomb interactions and short-range repulsion~\cite{Cohen1992,Posternak1994,Cochran1960}. PbTiO$_3$ later emerged as a key end member in relaxor ferroelectrics with giant electromechanical coupling~\cite{Park1997JAP}. At elevated pressures, increasingly complex behavior was reported, including pressure-induced morphotropic phase boundaries at cryogenic temperatures, where PbTiO$_3$ transitions from tetragonal $P4mm$ to monoclinic and then rhombohedral symmetry, enhancing the piezoelectric response~\cite{Wu2005,Ahart2008}. These studies led to proposals of reentrant ferroelectricity, in which the nonpolar $I4/mcm$ phase transforms into the polar $I4cm$ structure above $\sim$30~GPa~\cite{Kornev2005B}. A later study showed that room-temperature Raman and X-ray measurements confirmed the $P4mm \rightarrow I4/mcm$ transition near 18–20~GPa and suggested a recovery of a polar $I4cm$ state between 37–50~GPa based on improved fitting profiles~\cite{Janolin2008}. More recently, a combined theoretical and experimental investigation extended these studies to higher pressures and identified a centrosymmetric monoclinic $P2_1/m$ phase as the most stable structure of PbTiO$_3$ above $\sim$80~GPa~\cite{Cohen2024}. The centrosymmetric nature of this phase implies the absence of ferroelectricity and is consistent with second-harmonic generation measurements performed as a function of pressure, which ruled out any polar response at both 300~K and 10~K. This contrasting pressure response of ferroelectricity was highlighted in the classic work of Samara~\cite{Samara1996}, who demonstrated that pressure can either suppress or enhance ferroelectric instabilities depending on whether they originate from zone-center or zone-boundary soft modes. 

However, the stability and nature of PbTiO$_3$ under combined high $P$-$T$ conditions remain poorly constrained. Motivated by this unresolved question, we carried out high-pressure experiments on PbTiO$_3$ up to 100~GPa and 1400~K using laser heating. Such high–$P$–$T$ diamond-anvil cell (DAC) experiments have previously enabled the discovery and synthesis of numerous novel functional materials, including transition-metal nitrides, polar ordered oxynitride perovskites, and superconducting superhydrides~\cite{Young2006IrN2OsN2,Vadapoo2017Oxynitride,Somayazulu2019LaH10}. Until high pressure stability of perovskites was discovered, multicomponent oxides were thought to breakdown to simple oxides at high $P$–$T$ conditions. Contrary to theoretical predictions that did not account for the possible dissociation of PbTiO$_3$, our experiments show that the compound does not transform into the predicted monoclinic phase at high temperature but instead decomposes into PbO and TiO$_2$, while remaining in the $I4/mcm$ phase under room-temperature compression. Through PXRD-assisted crystal structure predictions with XtalOpt~\cite{Racioppi2025,Hajinazar2024}, we identify both the known $\alpha$-PbO phase and a previously unreported $\delta$-PbO polymorph. Together, these results reveal unexpected chemical reactivity and underscore the kinetic and thermodynamic complexity of PbTiO$_3$ under extreme pressure–temperature conditions.

Polycrystalline PbTiO$_3$ samples were studied using synchrotron X-ray diffraction in diamond anvil cells (DACs) at the European Synchrotron Radiation Facility (ESRF, beamline ID15B) and Advanced Photon Source (APS, beamline 16IDB), with selected experiments performed using laser heating. Density functional theory (DFT) calculations were carried out with VASP~\cite{Kresse1994} at the meta-GGA level using the r$^2$SCAN functional~\cite{Furness2020}. Further experimental and computational details are provided in the Supplemental Material.

\begin{figure}[htbp]

  \begin{overpic}[width=0.85\linewidth]{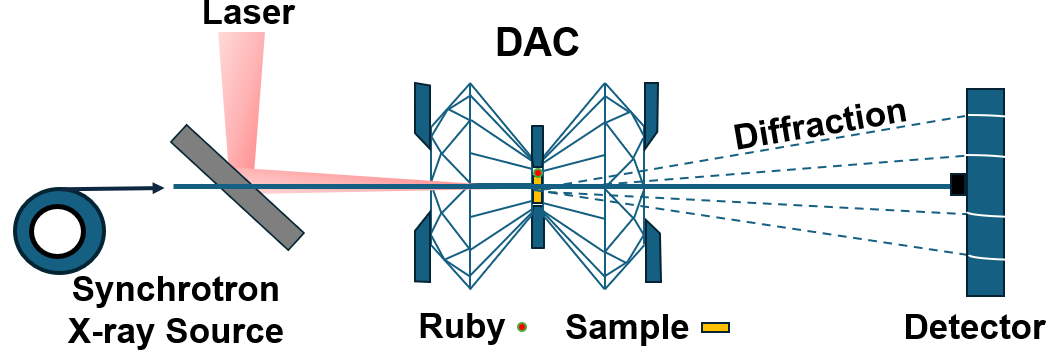}
    \put(2,25){\bfseries a}
  \end{overpic}

  \vspace{0.6cm}

  \begin{overpic}[width=0.85\linewidth]{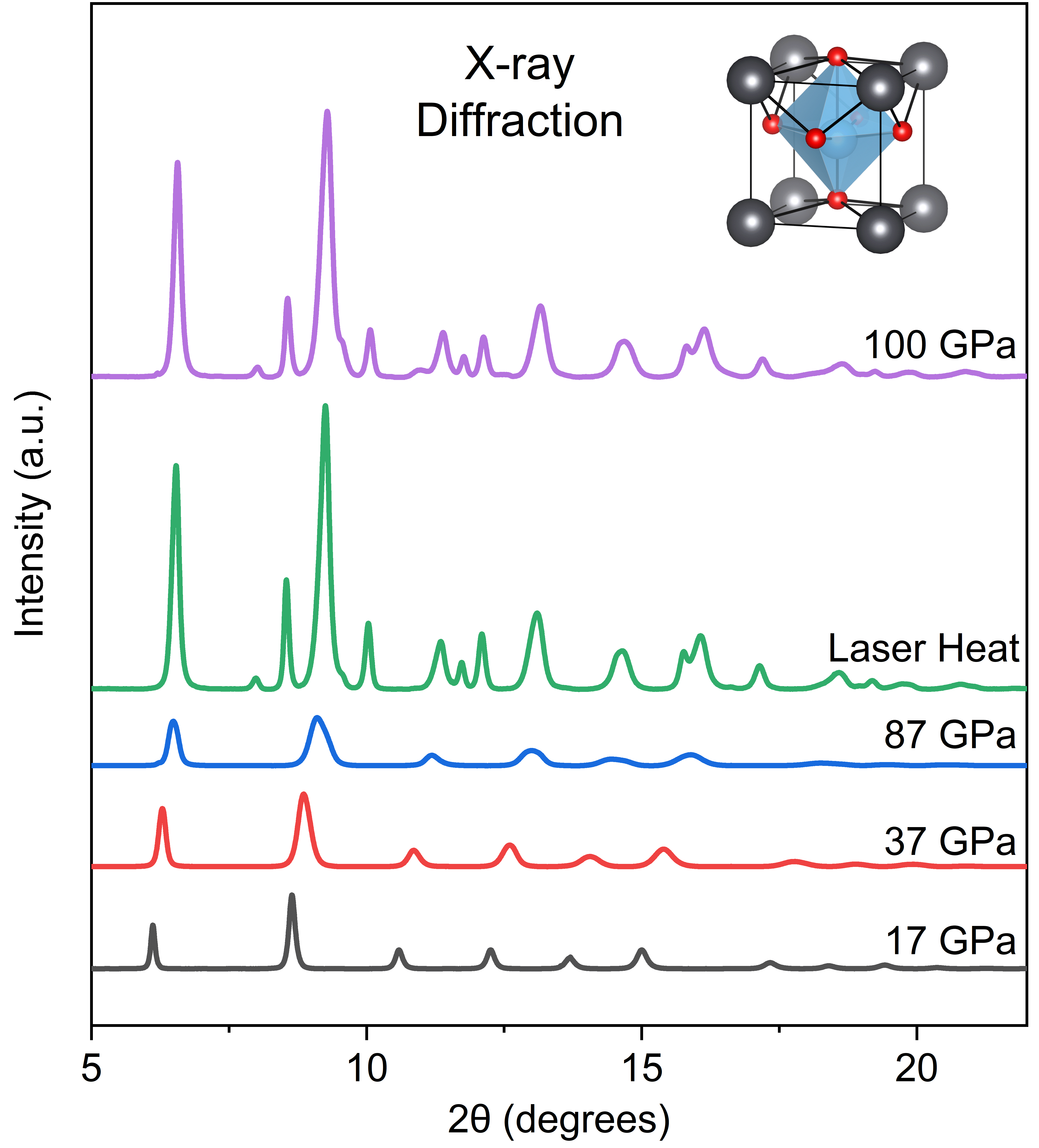}
    \put(9,96){\bfseries b}
  \end{overpic}

  \caption{(a) Schematic of the DAC laser-heating XRD experiment.
  (b) XRD patterns of PbTiO$_3$ collected from 17 to 100~GPa at 300~K, with laser heating performed at 87~GPa followed by additional compression. The crystal structure of tetragonal (\textit{P}4\textit{mm}) PbTiO$_3$ is shown within the plot for reference.}
  \label{fig:fig_1}
\end{figure}

High-pressure synchrotron X-ray diffraction (XRD) patterns of PbTiO$_3$ were collected from 17 to 100~GPa, with the experimental configuration and crystal structure shown in Fig.~\ref{fig:fig_1}. The sample was first laser-heated at 87~GPa and subsequently characterized by XRD upon quenching, rather than during in-situ heating. The diffraction patterns show systematic peak shifts consistent with progressive lattice compression up to 87~GPa. After heating, the post-quench XRD data exhibit substantial profile changes indicative of new structural products, and reflections from unreacted PbTiO$_3$ in $I4/mcm$ persist, consistent with one-sided laser heating that primarily heats the surface region.

\begin{figure}[htbp]

    \begin{overpic}[width=0.8\linewidth,trim=0 0 0 0,clip]{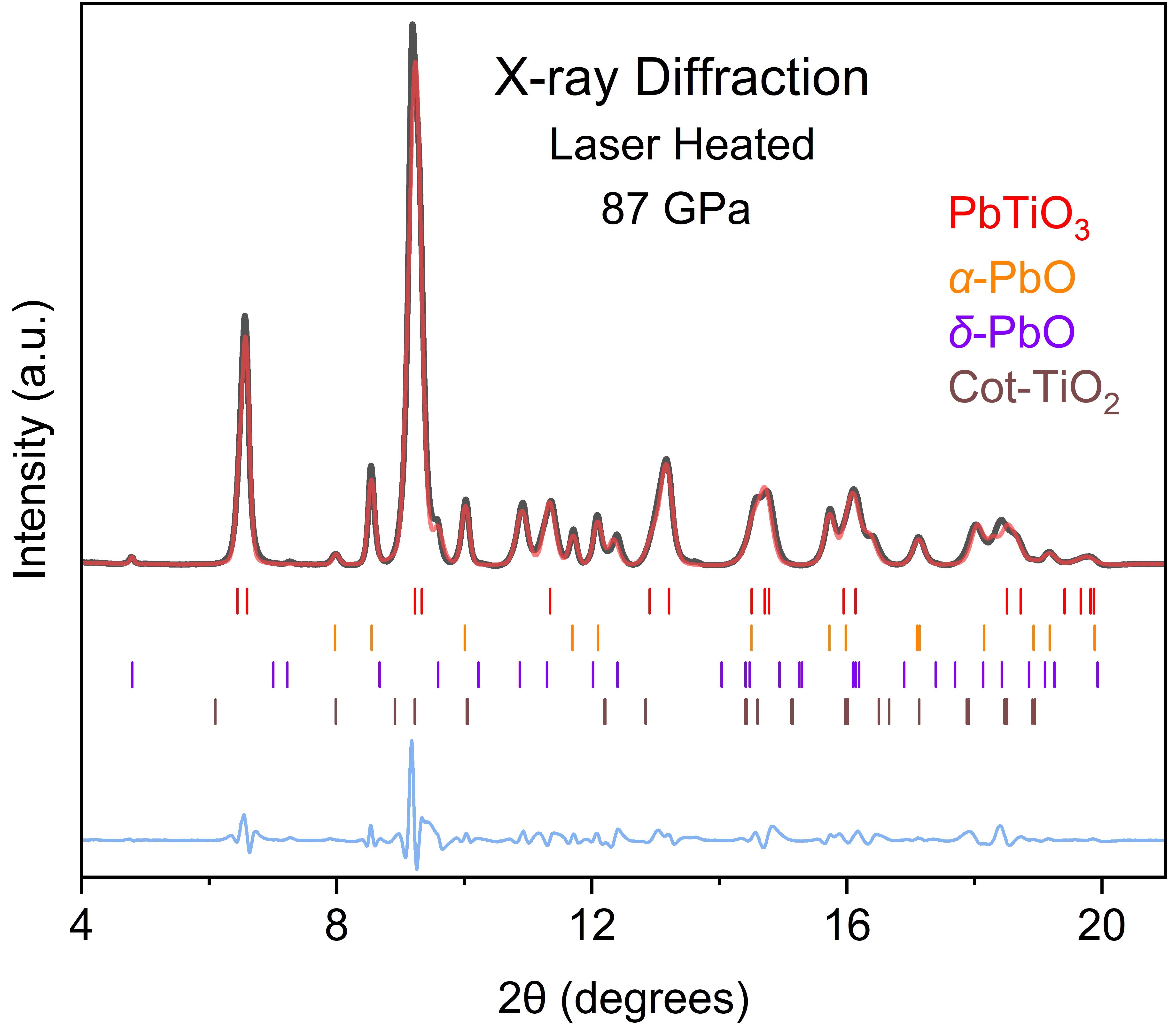}
        \put(8,83){\bfseries a}
    \end{overpic}
    \vspace{0.4cm}

    \begin{overpic}[width=0.8\linewidth,trim=0 0 0 0,clip]{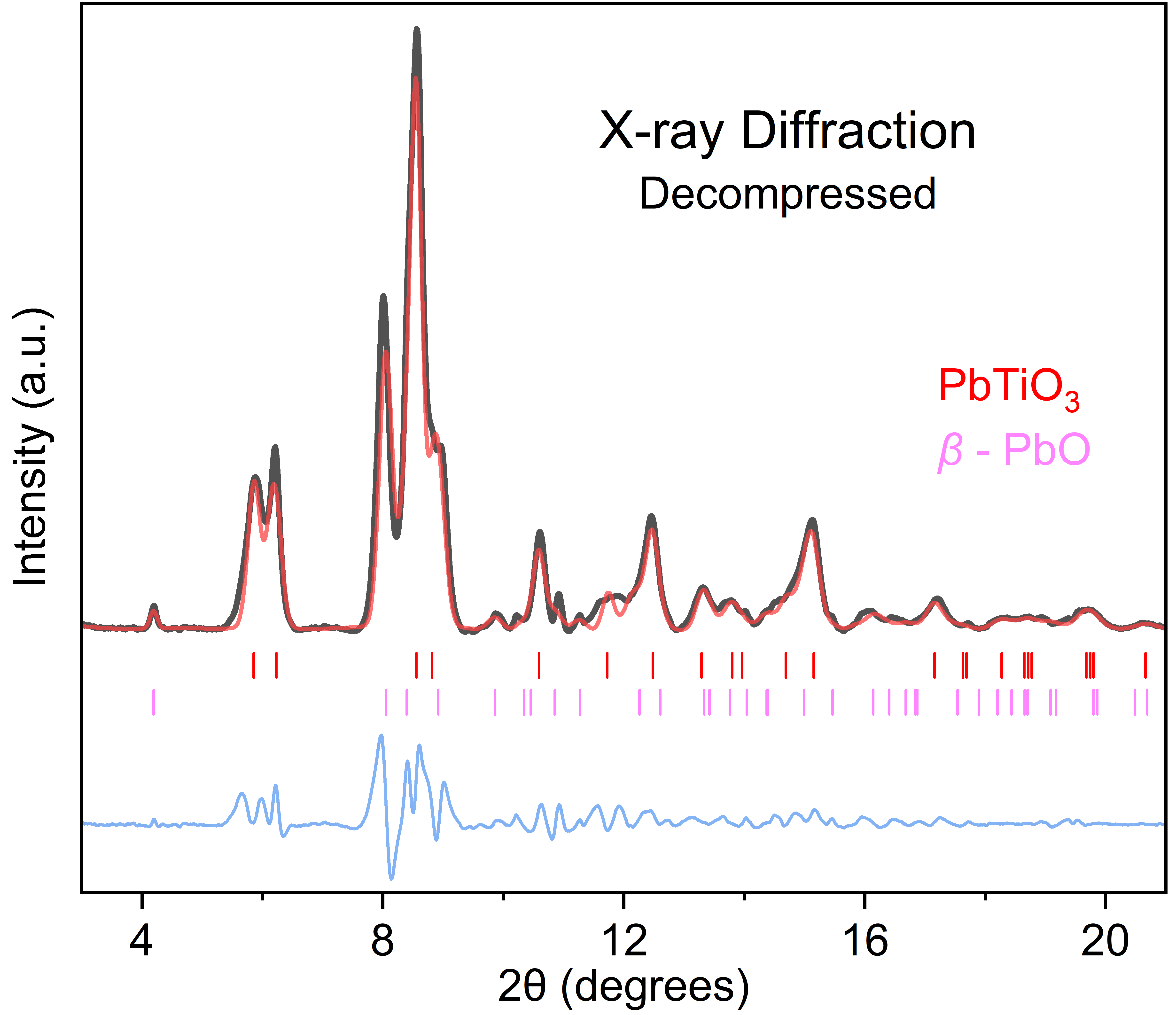}
        \put(8,82){\bfseries b}
    \end{overpic}

    \caption{(a) XRD pattern of a laser-heated PbTiO$_3$ sample at 87~GPa with Le Bail refinement. 
    (b) XRD pattern of the same sample after decompression to near-ambient pressure. 
    In both plots, the black, red, and blue lines represent the experimental data, fitted profile, and residuals, respectively. 
    Tick marks indicate Bragg reflection positions corresponding to PbTiO$_3$, PbO, and cotunnite-type TiO$_2$ phases.}

    \label{fig:fig_2}
\end{figure}

Figure \ref{fig:fig_2} shows the Le Bail refinement of the laser-heated sample at 87 GPa, revealing unreacted PbTiO$_3$ in the I4/mcm phase together with new high-pressure phases. These were found to have the same \textit{P}4/\textit{nmm} space group 
and identified as $\alpha$-PbO and $\delta$-PbO as discussed further below.
The refined lattice parameters are listed in Table~\ref{tab:lattice_hp}. PbO has been observed previously as $\alpha$-PbO (litharge, \textit{P}4/\textit{nmm}) and $\beta$-PbO (massicot, $Pbma$) at ambient pressure, and a shear-induced $\gamma$-PbO (\textit{Pm2$_1$n}) at 0.7--2.5~GPa~\cite{Adams1992,Giefers2007, Wang2013PbO}. Here, the first PbO product is a highly compressed form of $\alpha$-PbO with $c/a<1$ (ambient $c/a=1.26$~\cite{Adams1992,Giefers2007}), and the second \textit{P}4/\textit{nmm} form is a new phase, previously predicted for SnO~\cite{Nguyen2021}, which we term $\delta$-PbO. The structures are compared in detail below.

No diffraction peaks from TiO$_2$ were detected among the decomposition products. Although cotunnite-type TiO$_2$ is thermodynamically expected at these pressures~\cite{Dubrovinsky2001,AlKhatatbeh2009,Lee2022} and confirmed by our calculations, its absence likely reflects amorphization during the laser-heated decomposition of PbTiO$_3$, consistent with prior reports of pressure-induced amorphization in nanoscale TiO$_2$ polymorphs~\cite{Swamy2006,Wang2001,Li2010}. Figure~\ref{fig:fig_2}a shows the expected TiO$_2$ reflections overlapping with those of the PbO phases, which may obscure their detection. Kinetically hindered crystallization pathways can also lead to amorphous products, as observed for dense SiO$_2$ phases~\cite{Teter1998}. Upon decompression from 87~GPa, the unreacted PbTiO$_3$ reverted to its tetragonal (\textit{P}4\textit{mm}) structure, and both $\alpha$-PbO and $\delta$-PbO converted to $\beta$-PbO~\cite{Giefers2007,Sorrell1970thermal,Hill1985}. Refined lattice parameters are listed in Table~\ref{tab:lattice_decomp}.

Previous theoretical calculations, performed using the Wu–Cohen (WC) exchange–correlation functional, predicted a $P2_1/m$ post-perovskite transition of PbTiO$_3$ near 70~GPa~\cite{Cohen2024}, without considering decomposition. 
By explicitly including PbO and TiO$_2$ in the thermodynamic analysis, we find dissociation becomes favorable above $\sim$65~GPa. 
According to the reaction shown in Eq.~(\ref{eq:reaction1}), with the application of pressure and heat ($\Delta$):
\begin{equation}
\mathrm{PbTiO_3} \;\;\xrightarrow[\text{P $>$ 65 GPa}]{\Delta}\;\; \mathrm{PbO} + \mathrm{TiO_2}.
\label{eq:reaction1}
\end{equation}

The radial and axial temperature gradients of the single-sided laser heating~\cite{Meng2006,HPCAT2022,Garbarino2024,Fedotenko2019,Santoro2005,Shen1996} used here is known to produce hot spots and heterogeneity,~\cite{Kavner2008,Manga1996,Panero2001,Kavner2004}. As a result, we are able to observe multiple phases over a wide pressure range.  Notably, without heating, no dissociation was observed even above 65~GPa~\cite{Jabarova2011,AlZein2015,Janolin2008,Cohen2024,Kornev2005B}, underscoring the critical role of temperature in enabling diffusion and reaction.

\begin{figure}[htbp]
\begin{overpic}[width=0.8\linewidth]{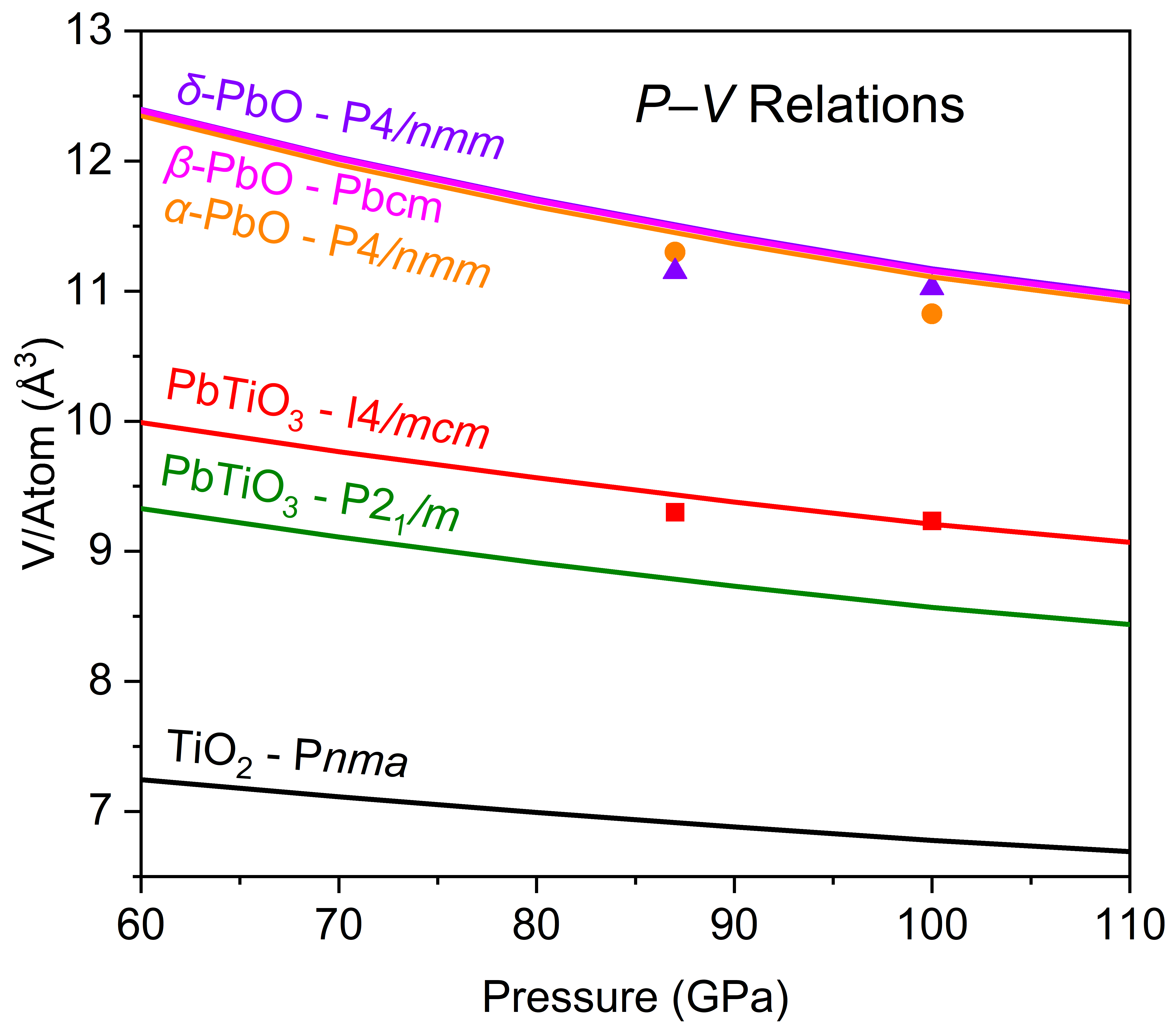}
\put(93,82){\bfseries a}
\end{overpic}
\vspace{0.4cm}
\begin{overpic}[width=0.8\linewidth]{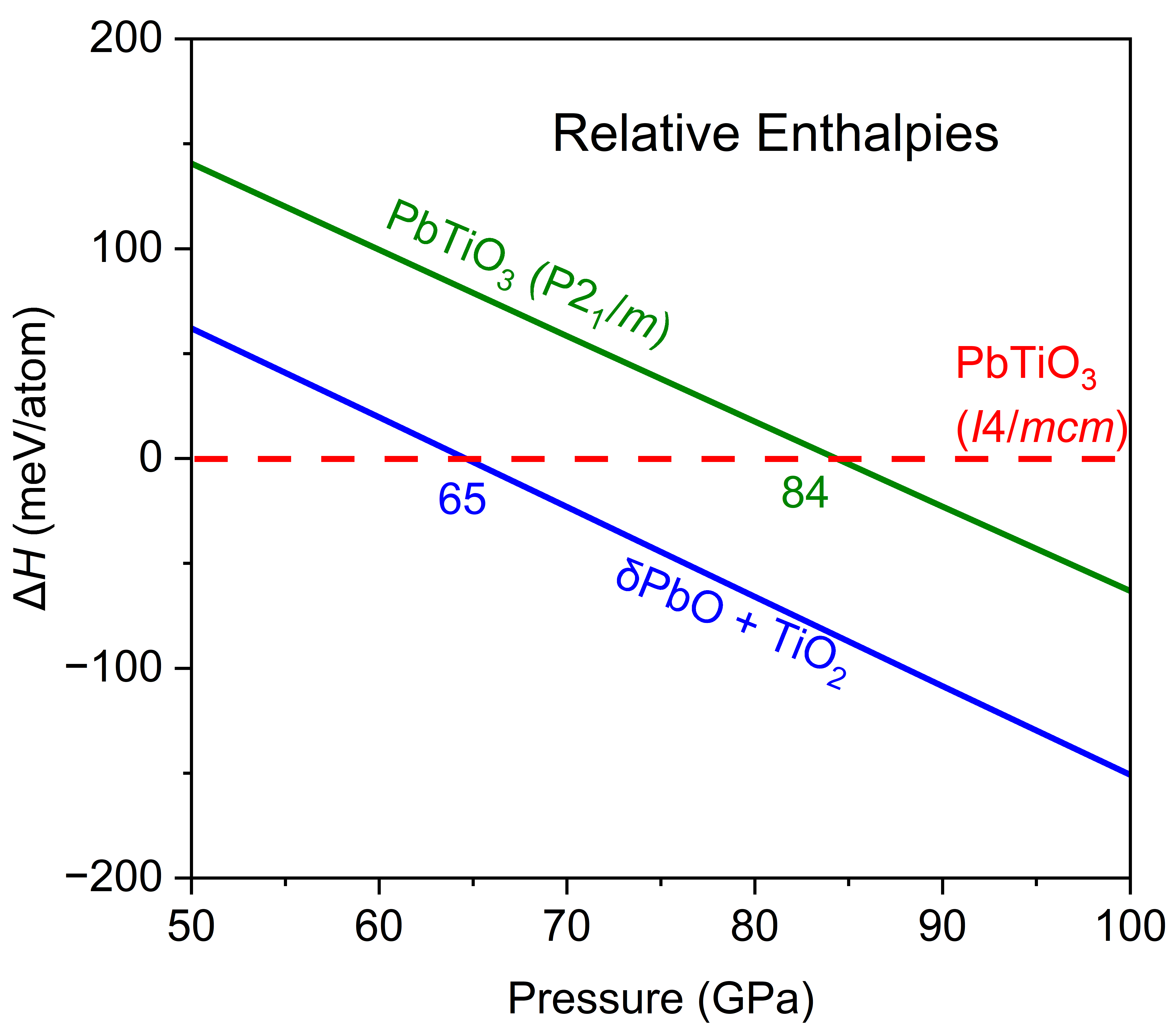}
\put(92,81){\bfseries b}
\end{overpic}
\caption{(a) $P$–$V$ relations for PbTiO$_3$ and its dissociation products. Curves represent DFT-calculated PbTiO$_3$ phases (\textit{I}4/\textit{mcm}, predicted \textit{P}2$_1$/\textit{m}), two high-pressure PbO phases (\textit{P}4/\textit{nmm}), and cotunnite-type TiO$_2$ (\textit{Pnma}). Symbols denote experimental data showing coexistence of PbTiO$_3$ with dissociation products. (b) Relative enthalpy as a function of pressure. The \textit{P}2$_1$/\textit{m} phase becomes marginally more stable near 84~GPa, and dissociation into $\delta$-PbO and TiO$_2$ becomes energetically favored above $\sim$65~GPa.}
\label{fig:fig_6_combined}
\end{figure}

Fig.~\ref{fig:fig_6_combined}a shows the pressure–volume ($P–V$) relations for PbTiO$_3$ and its dissociation products between 75 and 110~GPa. The experimental data points, representing unreacted PbTiO$_3$ and the coexisting PbO phases, fall close to the calculated equations of state. It is important to note that the three PbO polymorphs are very similar in density, consistent with their small relative volume differences. Fig.~\ref{fig:fig_6_combined}b displays the relative enthalpy curves for the competing PbTiO$_3$ phases and their dissociation products, showing that the \textit{I}4/\textit{mcm} structure is metastable above $\sim$65~GPa, where decomposition into $\delta$-PbO and cotunnite-type TiO$_2$ becomes thermodynamically favored over the previously predicted \textit{P}2$_1$/\textit{m} phase.

\begin{figure}[htbp]
    \includegraphics[width=0.95\linewidth]{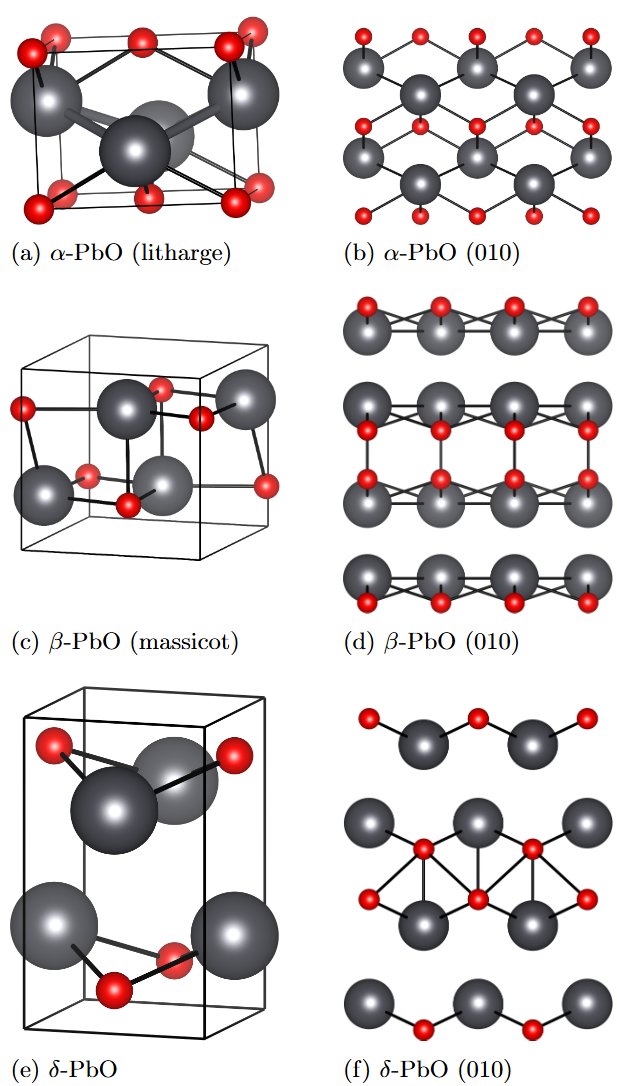}

  \caption{Crystal structures of the three PbO polymorphs.
    (a), (c), and (e) show the $\alpha$-, $\beta$-, and $\delta$-PbO unit cells, whereas 
    (b), (d), and (f) present the corresponding (010) projections that highlight their contrasting bonding topologies.
    $\alpha$-PbO forms a three-dimensional framework, 
    $\beta$-PbO exhibits puckered layers, 
    and $\delta$-PbO displays a two-dimensional layered structure.
    At 87~GPa, Pb–Pb distances are 2.871, 2.925, and 2.996~\AA{} for $\alpha$-, $\beta$-, and $\delta$-PbO, respectively.
    In $\alpha$-PbO, the Pb–Pb distance (2.871~\AA) represents the nearest interatomic spacing within its 3D network,
    whereas $\beta$-PbO and $\delta$-PbO possess interlayer separations of $\sim$3.5~\AA{} and $\sim$4.2~\AA{}, respectively, 
    between adjacent Pb–O layers.}
  \label{fig:PbO_structures}
\end{figure}

Three polymorphs of PbO were identified over the range of $P–T$ conditions explored (Fig.~\ref{fig:PbO_structures}): $\alpha$-PbO, $\beta$-PbO, and a new high-pressure $\delta$-PbO phase. The $\alpha$- and $\delta$-PbO polymorphs share the \textit{P}4/\textit{nmm} symmetry but differ in atomic arrangement: in both, Pb occupies the 2$c$ site, and O resides at 2$c$ ($z = 0.86$) for $\alpha$-PbO and 2$a$ for $\delta$-PbO, producing a three-dimensional framework in the former and a layered configuration in the latter (See Fig.~\ref{fig:PbO_structures}). In contrast, $\beta$-PbO (massicot, \textit{Pbcm}) adopts an orthorhombic structure with Pb and O on 4$d$ sites ($z_\mathrm{Pb} = 0.75$, $z_\mathrm{O} = 0.25$ at 50~GPa), forming puckered layers of corner- and edge-sharing PbO$_4$ pyramids in which O atoms exhibit distorted tetrahedral coordination by Pb ~\cite{Hill1985,Hill2025}. The diversity among these polymorphs reflects pressure-induced polytypism and the structural flexibility of PbO, reminiscent of NiAs-type FeO~\cite{Mazin1998}.

Direct experimental discrimination between metallic and semiconducting behavior in the high-pressure PbO polymorphs is not feasible due to the limited sample quantities, multiphase nature, and additional pressure-induced transformations of $\alpha$-PbO. Consequently, the electronic properties discussed above are assessed entirely through first-principles calculations.

\begin{figure}[htbp]
    \includegraphics[width=0.8\linewidth]{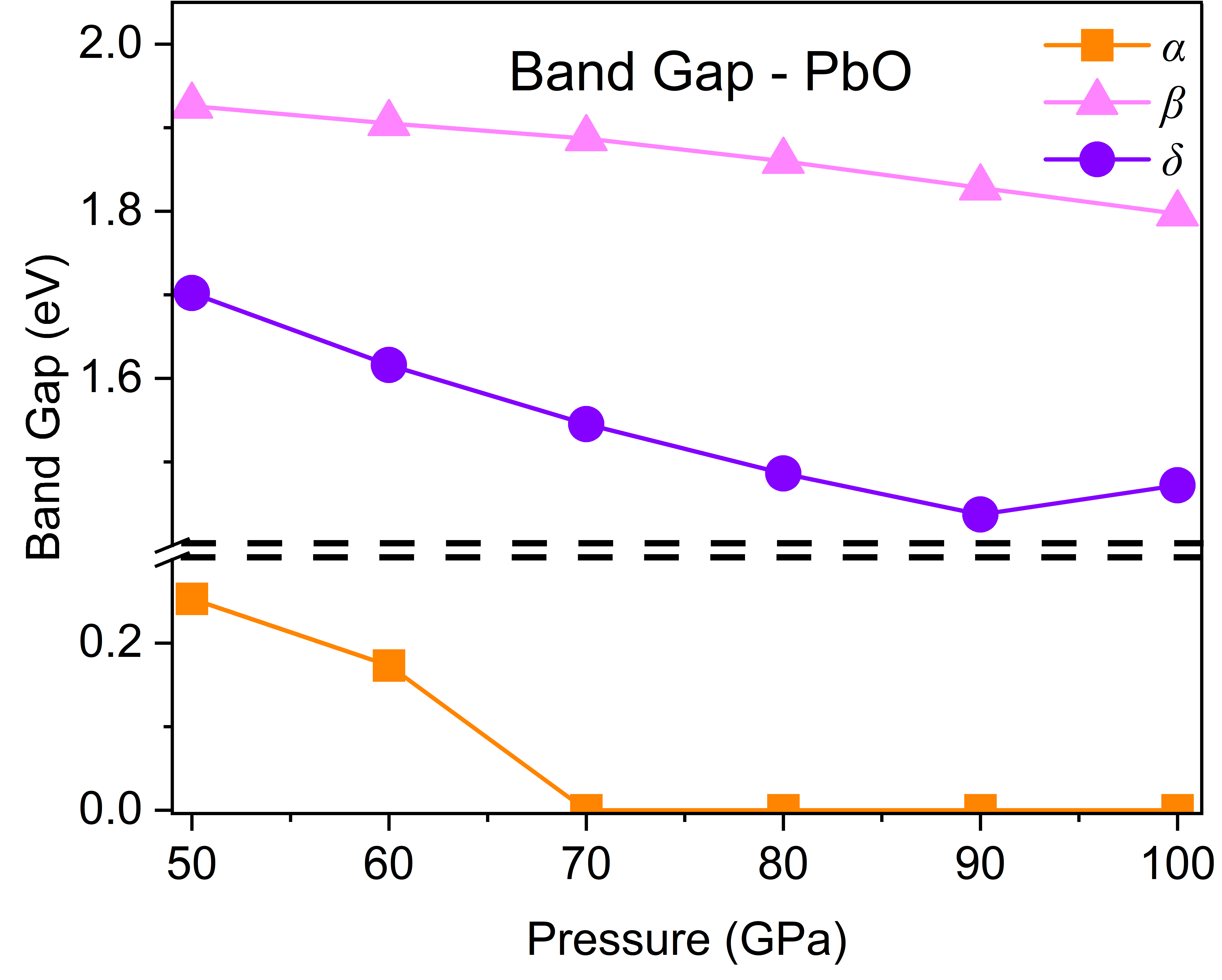}
    \caption{Band gaps of $\alpha$-, $\beta$- and $\delta$-PbO from 50–100~GPa, calculated with the DFT-r$^2$SCAN functional. $\alpha$-PbO metallizes near 70~GPa, whereas $\beta$- and $\delta$-PbO remain semiconducting.}
    \label{fig:fig_7}
\end{figure}

Electronic structure calculations reveal distinct pressure-dependent behaviors, as shown in Fig.~\ref{fig:fig_7}. $\alpha$-PbO undergoes band-gap closure and metallizes above $\sim$70~GPa, and $\beta$- and $\delta$-PbO remain semiconducting up to 100~GPa. The $\beta$ phase shows only a modest gap reduction from 1.9 to 1.8~eV between 50 and 100~GPa, whereas $\delta$-PbO retains a 1.4–1.7~eV gap~\cite{Lebeda2023}. The persistence of semiconducting behavior in $\beta$- and $\delta$-PbO reflects their larger Pb–Pb separations and layered frameworks, in contrast to the denser, more metallic $\alpha$-PbO network. In fact, it is likely that the Pb–Pb bonds play an important role in the incipient electronic properties of the PbO polymorphs. 

It is useful to contrast the behavior of PbTiO$_3$ with that of other oxide perovskites. Orthorhombic aluminate and stannate perovskites GdAlO$_3$, YAlO$_3$, and CaSnO$_3$ exhibit the well known octahedral tilting on compression~\cite{Ross2004_Gd,Ross2004_Y,Zhao2004_Ca}. KNbO$_3$, a prototypical ferroelectric perovskite, exhibits a sequence of pressure-induced phase transitions—from orthorhombic to tetragonal near 8~GPa, to cubic around 12~GPa, and finally to a distinct orthorhombic (GdFeO$_3$-type) structure near 40~GPa~\cite{Qi2023,Kobayashi2000}. In contrast, CaTiO$_3$, the mineral perovskite, remains stable on cold compression up to 60~GPa but transforms upon heating at higher  pressure to a monoclinic CaTi$_2$O$_5$ phase and CaO~\cite{Truffet2023,Guennou2010}. Of the magnetic perovskites, LaCrO$_3$ undergoes a transition from rhombohedral to orthorhombic near 5~GPa~\cite{Zhou2011,Hashimoto1998}, whereas HoCrO$_3$ transforms to a monoclinic phase above 20~GPa~\cite{Bhadram2014,Mall2023}. Although there is no evidence of PbTiO$_3$ in the deep mantle, its behavior may be compared to that of the well studied (Mg,Fe)SiO$_3$. Iron-rich (Mg$_x$Fe$_{1-x}$)SiO$_3$ perovskite breaks down to form mixed oxides at lower mantle $P$-$T$ conditions similar to those explored here~\cite{Hemley1992_SilicatePerovskite}, whereas magnesium-rich (Mg$_x$Fe$_{1-x}$)SiO$_3$ remains stable but transforms to a post-perovskite phase at higher pressures~\cite{Hirose2006}.

In conclusion, PbTiO$_3$ exhibits a remarkable temperature-dependent response to compression. Two metastable pathways can be accessed on compression: one leads toward post-perovskite structures, whereas another preserves a nonpolar perovskite configuration that remains stable up to 100~GPa at room temperature ~\cite{Cohen2024}. At elevated $P$-$T$ conditions, however, the system follows an equilibrium pathway that drives disproportionation into PbO and TiO$_2$, producing a novel $\delta-$PbO polymorph distinct from the well-known $\alpha$-PbO (litharge) and $\beta$-PbO (massicot) structures. This new PbO polymorph is predicted to remain semiconducting to at least 100~GPa. These results illustrate how temperature can be used to control kinetic pathways for the synthesis of novel phases at high pressures: we demonstrate that in addition to the predicted reentrant ferroelectric and post-perovskite phases, PbTiO$_3$ undergoes disproportionation, thereby expanding our understanding of the chemistry and diverse properties of oxide perovskites. Finally, we point out that early high-pressure studies suggested that all complex oxides ultimately disproportionate to form simple close packed oxide phases at sufficiently high pressures~\cite{Birch1952Elasticity}. Later, static compression experiments demonstrated that a wide range of complex oxides instead stabilize dense crystal structures, notably perovskite~\cite{Hemley1992_SilicatePerovskite} and later post-perovskite~\cite{Hirose2006} phases.  Our findings reveal a pathway for PbTiO$_3$ that follows the original paradigm.

\section*{Acknowledgments} This work was supported by the U.S. National Science Foundation (NSF) under Grants DMR-2119308 (HF, MA, RJH), DMR-2119065 (EZ), and PHY-2020249 (CMAP; SR). Additional support was provided by the U.S. Department of Energy (DOE), National Nuclear Security Administration (NNSA), through the Chicago/DOE Alliance Center (CDAC) under Award DE-NA0004153 (HF, MA, EZ, RJ), and by the DOE Office of Fusion Energy Sciences (OFES) under Award DE-SC0020340 (EZ, RH). We acknowledged the European Synchrotron Radiation Facility (ESRF) for provision of beamtime under proposal 5917. Portions of this work were performed at HPCAT (Sector 16), Advanced Photon Source (APS), Argonne National Laboratory (ANL), a DOE Office of Science (SC) user facility operated under Contract DE-AC02-06CH11357. HPCAT operations were supported by the DOE-NNSA Office of Experimental Sciences. Calculations were performed at the Center for Computational Research, University at Buffalo (http://hdl.handle.net/10477/79221). This work is a contribution of the George W. Crabtree Institute for Discovery and Sustainability.

\RefHeading
\bibliographystyle{apsrev4-2}
\bibliography{PTO}

@PREAMBLE{
 "\providecommand{\noopsort}[1]{}" 
 # "\providecommand{\singleletter}[1]{#1}%" 
}

@article{Nguyen2021,
  author  = {L. T. Nguyen and G. Makov},
  title   = {High-pressure phases of {SnO} and {PbO}: A density functional theory combined with an evolutionary algorithm approach},
  journal = {Materials},
  volume  = {14},
  pages   = {6552},
  year    = {2021},
  doi     = {10.3390/ma14216552}
}

@article{Manga1996,
  author  = {M. Manga and R. Jeanloz},
  title   = {Thermal conductivity and temperature gradients in the laser-heated diamond anvil cell},
  journal = {Geophys. Res. Lett.},
  volume  = {23},
  number  = {14},
  pages   = {1845--1848},
  year    = {1996},
  doi     = {10.1029/96GL01602}
}

@article{Panero2001,
  author  = {W. R. Panero and R. Jeanloz},
  title   = {Temperature gradients in the laser-heated diamond anvil cell},
  journal = {J. Geophys. Res. Solid Earth},
  volume  = {106},
  number  = {B4},
  pages   = {6493--6498},
  year    = {2001},
  doi     = {10.1029/2000JB900423}
}

@article{Kavner2004,
  author  = {A. Kavner and W. R. Panero},
  title   = {Temperature gradients and evaluation of thermoelastic properties in the synchrotron-based laser-heated diamond cell},
  journal = {Phys. Earth Planet. Inter.},
  volume  = {143-144},
  pages   = {527--539},
  year    = {2004},
  doi     = {10.1016/j.pepi.2003.12.016}
}

@article{Kavner2008,
  author  = {A. Kavner and C. Nugent},
  title   = {Precise measurements of radial temperature gradients in the laser-heated diamond anvil cell},
  journal = {Rev. Sci. Instrum.},
  volume  = {79},
  number  = {2},
  pages   = {024902},
  year    = {2008},
  doi     = {10.1063/1.2841173}
}

@article{Lebeda2023,
  author  = {T. Lebeda and T. Aschebrock and J. Sun and L. Leppert and S. Kümmel},
  title   = {Right band gaps for the right reason at low computational cost with a meta-{GGA}},
  journal = {Phys. Rev. Mater.},
  volume  = {7},
  pages   = {093803},
  year    = {2023},
  doi     = {10.1103/PhysRevMaterials.7.093803}
}

@article{Zhang2023advances,
  author  = {L. Zhang and L. Mei and K. Wang and Y. Lv and S. Zhang and Y. Lian and X. Liu and Z. Ma and G. Xiao and Q. Liu and S. Zhai and S. Zhang and G. Liu and L. Yuan and B. Guo and Z. Chen and K. Wei and A. Liu and S. Yue and G. Niu and X. Pan and J. Sun and Y. Hua and W.-Q. Wu and D. Di and B. Zhao and J. Tian and Z. Wang and Y. Yang and L. Chu and M. Yuan and H. Zeng and H.-L. Yip and K. Yan and W. Xu and L. Zhu and W. Zhang and G. Xing and F. Gao and L. Ding},
  title   = {Advances in the application of perovskite materials},
  journal = {Nano-Micro Lett.},
  volume  = {15},
  pages   = {177},
  year    = {2023},
  doi     = {10.1007/s40820-023-01140-3}
}

@article{Baudry2017,
  author  = {L. Baudry and I. Lukyanchuk and V. M. Vinokur},
  title   = {Ferroelectric symmetry-protected multibit memory cell},
  journal = {Sci. Rep.},
  volume  = {7},
  pages   = {42196},
  year    = {2017},
  doi     = {10.1038/srep42196}
}

@article{Shin2018,
  author  = {H. W. Shin and J. Y. Son},
  title   = {Nonvolatile ferroelectric memory based on {PbTiO}$_3$ gated single-layer {MoS}$_2$ field-effect transistor},
  journal = {Electron. Mater. Lett.},
  volume  = {14},
  pages   = {59--63},
  year    = {2018},
  doi     = {10.1007/s13391-017-7137-y}
}

@article{Moret2002,
  author  = {M. P. Moret and M. A. C. Devillers and K. W{\"o}rhoff and P. K. Larsen},
  title   = {Optical properties of {PbTiO}$_3$, {PbZr}$_x${Ti}$_{1-x}${O}$_3$, and {PbZrO}$_3$ films deposited by metalorganic chemical vapor on {SrTiO}$_3$},
  journal = {J. Appl. Phys.},
  volume  = {92},
  number  = {1},
  pages   = {468--474},
  year    = {2002},
  doi     = {10.1063/1.1486048}
}

@article{Yoon1993,
  author  = {Y. S. Yoon and W. N. Kang and S. S. Yom and T. W. Kim and M. Jung and H. J. Kim and T. H. Park and H. K. Na},
  title   = {Electrical and optical properties of {PbTiO}$_3$ thin films on p-{Si} grown by metalorganic chemical vapor deposition at low temperature},
  journal = {Appl. Phys. Lett.},
  volume  = {63},
  number  = {8},
  pages   = {1104--1106},
  year    = {1993},
  doi     = {10.1063/1.109794}
}

@article{Kim1995,
  author  = {Y. T. Kim and C. W. Lee},
  title   = {Dielectric properties of {PbTiO}$_3$ thin film capacitors deposited on tungsten nitride/tungsten bilayers},
  journal = {Ferroelectrics},
  volume  = {166},
  number  = {1},
  pages   = {159--163},
  year    = {1995},
  doi     = {10.1080/00150199508223584}
}

@article{Zhang2013,
  author  = {S. Zhang and F. Li and J. Luo and R. Sahul and T. R. Shrout},
  title   = {Relaxor-{PbTiO}$_3$ single crystals for various applications},
  journal = {IEEE Trans. Ultrason. Ferroelectr. Freq. Control},
  volume  = {60},
  number  = {8},
  pages   = {1572--1580},
  year    = {2013},
  doi     = {10.1109/TUFFC.2013.2737}
}

@article{Kanda2012,
  author  = {K. Kanda and J. Inoue and T. Saito and T. Fujita and K. Higuchi and K. Maenaka},
  title   = {Fabrication and characterization of double-layer {Pb}({Zr},{Ti}){O}$_3$ thin films for micro-electromechanical systems},
  journal = {Jpn. J. Appl. Phys.},
  volume  = {51},
  number  = {9S1},
  pages   = {09LD12},
  year    = {2012},
  doi     = {10.1143/JJAP.51.09LD12}
}

@article{Okuyama1991,
  author  = {M. Okuyama and Y. Hamakawa},
  title   = {{PbTiO}$_3$ ferroelectric thin films and their pyroelectric application},
  journal = {Ferroelectrics},
  volume  = {118},
  number  = {1},
  pages   = {261--278},
  year    = {1991},
  doi     = {10.1080/00150199108014765}
}

@article{Cohen1990,
  author  = {R. E. Cohen and H. Krakauer},
  title   = {Lattice dynamics and origin of ferroelectricity in {BaTiO}$_3$: Linearized-augmented-plane-wave total-energy calculations},
  journal = {Phys. Rev. B},
  volume  = {42},
  number  = {10},
  pages   = {6416--6423},
  year    = {1990},
  doi     = {10.1103/PhysRevB.42.6416}
}

@article{Cohen1992,
  author  = {R. E. Cohen},
  title   = {Origin of ferroelectricity in perovskite oxides},
  journal = {Nature},
  volume  = {358},
  pages   = {136--138},
  year    = {1992},
  doi     = {10.1038/358136a0}
}

@article{Yoshiasa2016,
  author  = {A. Yoshiasa and T. Nakatani and A. Nakatsuka and M. Okube and K. Sugiyama and T. Mashimo},
  title   = {High-temperature single-crystal x-ray diffraction study of tetragonal and cubic perovskite-type {PbTiO}$_3$ phases},
  journal = {Acta Crystallogr. B},
  volume  = {72},
  number  = {3},
  pages   = {381--388},
  year    = {2016},
  doi     = {10.1107/S2052520616005114}
}

@article{Shirane1955,
  author  = {Shirane, G. and Pepinsky, R. and Frazer, B. C.},
  title   = {X-Ray and Neutron Diffraction Study of Ferroelectric {PbTiO}$_3$},
  journal = {Phys. Rev.},
  volume  = {97},
  number  = {4},
  pages   = {1179--1180},
  year    = {1955},
  doi     = {10.1103/PhysRev.97.1179}
}

@inproceedings{Peng1991,
  author    = {C. H. Peng and J.-F. Chang and S. B. Desu},
  title     = {Optical properties of {PZT}, {PLZT}, and {PNZT} thin films},
  booktitle = {MRS {P}roc.},
  volume    = {243},
  pages     = {21--26},
  year      = {1991},
  doi       = {10.1557/PROC-243-21}
}

@article{Sanjurjo1983,
  author  = {J. A. Sanjurjo and E. L{\'o}pez-Cruz and G. Burns},
  title   = {High-pressure Raman study of zone-center phonons in {PbTiO}$_3$},
  journal = {Phys. Rev. B},
  volume  = {28},
  pages   = {7260--7268},
  year    = {1983},
  doi     = {10.1103/PhysRevB.28.7260}
}

@article{Wu2005,
  author  = {Z. Wu and R. E. Cohen},
  title   = {Pressure-induced anomalous phase transitions and colossal enhancement of piezoelectricity in {PbTiO}$_3$},
  journal = {Phys. Rev. Lett.},
  volume  = {95},
  number  = {3},
  pages   = {037601},
  year    = {2005},
  doi     = {10.1103/PhysRevLett.95.037601}
}

@article{Ahart2008,
  author  = {M. Ahart and M. Somayazulu and R. E. Cohen and P. Ganesh and P. Dera and H. K. Mao and R. J. Hemley and Y. Ren and P. Liermann and Z. Wu},
  title   = {Origin of morphotropic phase boundaries in ferroelectrics},
  journal = {Nature},
  volume  = {451},
  pages   = {545--548},
  year    = {2008},
  doi     = {10.1038/nature06459}
}

@article{Cohen2024,
  author  = {R. E. Cohen and Y. Lin and M. Ahart and R. J. Hemley},
  title   = {Absence of high-pressure ground-state reentrant ferroelectricity in {PbTiO}$_3$},
  journal = {Phys. Rev. Lett.},
  volume  = {133},
  number  = {23},
  pages   = {236801},
  year    = {2024},
  doi     = {10.1103/PhysRevLett.133.236801}
}

@inproceedings{Shen1996,
  author    = {G. Shen and H. K. Mao and R. J. Hemley},
  title     = {{Laser{-}Heating} Diamond{-}Anvil Cell Technique: Double{-}Sided Heating with Multimode {N}d:{YAG} Laser},
  booktitle = {{Advanced Materials} '96: {New Trends in High{-}Pressure Research}},
  series    = {Proc. 3rd NIRIM Int. Symp. Advanced Materials (ISAM '96)},
  address   = {Tsukuba, Japan},
  pages     = {149--152},
  year      = {1996},
  publisher = {NIRIM}
}

@incollection{Santoro2005,
  author    = {M. Santoro and J.-F. Lin and V. V. Struzhkin and H.-K. Mao and R. J. Hemley},
  title     = {In situ Raman spectroscopy with laser-heated diamond anvil cells},
  booktitle = {Advances in {H}igh-{P}ressure {T}echnology for {G}eophysical {A}pplications},
  editor    = {J. Chen and Y. Wang and T. S. Duffy and G. Shen and L. F. Dobrzhinetskaya},
  publisher = {Elsevier},
  year      = {2005},
  chapter   = {20},
  pages     = {413--423},
  isbn      = {9780444519795},
  doi       = {10.1016/B978-044451979-5.50022-3}
}

@article{Janolin2008,
  author  = {P.-E. Janolin and P. Bouvier and J. Kreisel and P. A. Thomas and I. A. Kornev and L. Bellaiche and W. Crichton and M. Hanfland and B. Dkhil},
  title   = {High-pressure effect on {PbTiO}$_3$: An investigation by Raman and x-ray scattering up to 63 {GP}a},
  journal = {Phys. Rev. Lett.},
  volume  = {101},
  number  = {23},
  pages   = {237601},
  year    = {2008},
  doi     = {10.1103/PhysRevLett.101.237601}
}

@article{AlZein2015,
  author  = {A. Al-Zein and P. Bouvier and A. Kania and C. Levelut and B. Hehlen and V. Nassif and T. C. Hansen and P. Fertey and J. Haines and J. Rouquette},
  title   = {High pressure single crystal x-ray and neutron powder diffraction study of the ferroelectric–paraelectric phase transition in {PbTiO}$_3$},
  journal = {J. Phys. D: Appl. Phys.},
  volume  = {48},
  number  = {50},
  pages   = {504008},
  year    = {2015},
  doi     = {10.1088/0022-3727/48/50/504008}
}

@article{Jabarova2011,
  author  = {S. G. Jabarova and D. P. Kozlenko and S. E. Kichanova and A. V. Belushkin and B. N. Savenko and R. Z. Mextieva and C. Lathe},
  title   = {High-pressure effect on the ferroelectric–paraelectric transition in {PbTiO}$_3$},
  journal = {Phys. Solid State},
  volume  = {53},
  pages   = {2300--2304},
  year    = {2011},
  doi     = {10.1134/S1063783411110102}
}

@article{Guennou2010,
  author  = {M. Guennou and P. Bouvier and B. Krikler and J. Kreisel and R. Haumont and G. Garbarino},
  title   = {High-pressure investigation of {CaTiO}$_3$ up to 60 {GP}a using x-ray diffraction and Raman spectroscopy},
  journal = {Phys. Rev. B},
  volume  = {82},
  number  = {13},
  pages   = {134101},
  year    = {2010},
  doi     = {10.1103/PhysRevB.82.134101}
}

@article{Truffet2023,
  author  = {B. Truffet and G. Fiquet and G. Morard and M. A. Baron and F. Miozzi and M. Harmand and A. Ravasio and M. Mezouar and F. Guyot},
  title   = {High pressure dissociation of {CaTiO}$_3$ perovskite into {CaO} and {CaTi}$_2${O}$_5$},
  journal = {Phys. Earth Planet. Inter.},
  volume  = {334},
  pages   = {106968},
  year    = {2023},
  doi     = {10.1016/j.pepi.2022.106968}
}

@article{Mall2023,
  author  = {A. K. Mall and N. Garg and A. K. Verma and D. Errandonea and A. V. Chitnis and V. Srihari and R. Gupta},
  title   = {Discovery of high-pressure post-perovskite phase in {HoCrO}$_3$},
  journal = {J. Phys. Chem. Solids},
  volume  = {172},
  pages   = {111078},
  year    = {2023},
  doi     = {10.1016/j.jpcs.2022.111078}
}

@article{Bhadram2014,
  author  = {V. S. Bhadram and D. Swain and R. Dhanya and M. Polentarutti and A. Sundaresan and C. Narayana},
  title   = {Effect of pressure on octahedral distortions in {RCrO}$_3$ ({R = Lu, Tb, Gd, Eu, Sm}): The role of {R}-ion size and its implications},
  journal = {Mater. Res. Express},
  volume  = {1},
  number  = {2},
  pages   = {026111},
  year    = {2014},
  doi     = {10.1088/2053-1591/1/2/026111}
}

@article{Zhou2011,
  author  = {J.-S. Zhou and J. A. Alonso and A. Muñoz and M. T. Fernández-Díaz and J. B. Goodenough},
  title   = {Magnetic structure of {LaCrO}$_3$ perovskite under high pressure from in situ neutron diffraction},
  journal = {Phys. Rev. Lett.},
  volume  = {106},
  pages   = {057201},
  year    = {2011},
  doi     = {10.1103/PhysRevLett.106.057201}
}

@article{Hashimoto1998,
  author  = {T. Hashimoto and N. Matsushita and Y. Murakami and N. Kojima and K. Yoshida and H. Tagawa and M. Dokiya and T. Kikegawa},
  title   = {Pressure-induced structural phase transition of {LaCrO}$_3$},
  journal = {Solid State Commun.},
  volume  = {108},
  number  = {9},
  pages   = {691--694},
  year    = {1998},
  doi     = {10.1016/S0038-1098(98)00446-3}
}

@article{Kobayashi2000,
  author  = {Y. Kobayashi and S. Endo and T. Ashida and L. C. Ming and T. Kikegawa},
  title   = {High-pressure phase above 40 {GP}a in ferroelectric {KNbO}$_3$},
  journal = {Phys. Rev. B},
  volume  = {61},
  number  = {9},
  pages   = {5819--5822},
  year    = {2000},
  doi     = {10.1103/PhysRevB.61.5819}
}

@article{Ross2004_Gd,
  author  = {N. L. Ross and J. Zhao and R. J. Angel},
  title   = {High-pressure structural behavior of {Gd}{Al}{O}$_3$ and {Gd}{Fe}{O}$_3$ perovskites},
  journal = {J. Solid State Chem.},
  volume  = {177},
  number  = {10},
  pages   = {3768--3775},
  year    = {2004},
  doi     = {10.1016/j.jssc.2004.07.002}
}

@article{Ross2004_Y,
  author  = {N. L. Ross and J. Zhao and R. J. Angel},
  title   = {High-pressure single-crystal X-ray diffraction study of {Y}{Al}{O}$_3$ perovskite},
  journal = {J. Solid State Chem.},
  volume  = {177},
  number  = {4-5},
  pages   = {1276--1284},
  year    = {2004},
  doi     = {10.1016/j.jssc.2003.11.014}
}

@article{Zhao2004_Ca,
  author  = {J. Zhao and N. L. Ross and R. J. Angel},
  title   = {Tilting and distortion of {CaSnO}$_3$ perovskite to 7 {GP}a determined from single-crystal X-ray diffraction},
  journal = {Phys. Chem. Miner.},
  volume  = {31},
  pages   = {299--305},
  year    = {2004},
  doi     = {10.1007/s00269-004-0391-1}
}

@article{Kresse1994,
  author  = {G. Kresse and J. Hafner},
  title   = {Ab initio molecular-dynamics simulation of the liquid-metal--amorphous-semiconductor transition in germanium},
  journal = {Phys. Rev. B},
  volume  = {49},
  number  = {20},
  pages   = {14251--14269},
  year    = {1994},
  doi     = {10.1103/PhysRevB.49.14251}
}

@article{Furness2020,
  author  = {J. W. Furness and A. D. Kaplan and J. Ning and J. P. Perdew and J. Sun},
  title   = {Accurate and numerically efficient r$^2$SCAN meta-generalized gradient approximation},
  journal = {J. Phys. Chem. Lett.},
  volume  = {11},
  pages   = {8208--8215},
  year    = {2020},
  doi     = {10.1021/acs.jpclett.0c02405}
}

@article{Blochl1994,
  author  = {P. E. Bl{\"o}chl},
  title   = {Projector augmented-wave method},
  journal = {Phys. Rev. B},
  volume  = {50},
  number  = {24},
  pages   = {17953--17979},
  year    = {1994},
  doi     = {10.1103/PhysRevB.50.17953}
}

@article{Monkhorst1976,
  author    = {H. J. Monkhorst and J. D. Pack},
  title     = {Special points for Brillouin-zone integrations},
  journal   = {Phys. Rev. B},
  volume    = {13},
  pages     = {5188--5192},
  year      = {1976},
  doi       = {10.1103/PhysRevB.13.5188}
}

@article{Blochl1994_tetra,
  author    = {P. E. Bl{\"o}chl and O. Jepsen and O. K. Andersen},
  title     = {Improved tetrahedron method for Brillouin-zone integrations},
  journal   = {Phys. Rev. B},
  volume    = {49},
  pages     = {16223--16233},
  year      = {1994},
  doi       = {10.1103/PhysRevB.49.16223}
}

@article{Togo2008,
  author    = {A. Togo and F. Oba and I. Tanaka},
  title     = {First-principles calculations of the ferroelastic transition between rutile-type and {CaCl}$_2$-type {SiO}$_2$ at high pressures},
  journal   = {Phys. Rev. B},
  volume    = {78},
  pages     = {134106},
  year      = {2008},
  doi       = {10.1103/PhysRevB.78.134106}
}

@article{HPCAT2022,
  author  = {A. Bommannavar and P. Chow and R. Ferry and R. Hrubiak and F. Humble and C. Kenney-Benson and M. Ly and Y. Meng and C. Park and D. Popov and E. Rod and M. Somayazulu and G. Shen and D. Smith and J. Smith and Y. Xiao and N. Velisavljevic},
  title   = {Overview of {HPCAT} and capabilities for studying minerals and various other materials at high-pressure conditions},
  journal = {Phys. Chem. Miner.},
  volume  = {49},
  number  = {9},
  pages   = {36},
  year    = {2022},
  doi     = {10.1007/s00269-022-01209-2}
}

@article{Meng2006,
  author  = {Y. Meng and G. Shen and H. K. Mao},
  title   = {Double-sided laser heating system at {HPCAT} for in situ x-ray diffraction at high pressures and high temperatures},
  journal = {J. Phys.: Condens. Matter},
  volume  = {18},
  number  = {25},
  pages   = {S1097--S1103},
  year    = {2006},
  doi     = {10.1088/0953-8984/18/25/S17}
}

@article{Fedotenko2019,
  author  = {T. Fedotenko and L. Dubrovinsky and G. Aprilis and E. Koemets and A. Snigirev and I. Snigireva and A. Barannikov and P. Ershov and F. Cova and M. Hanfland and N. Dubrovinskaia},
  title   = {Laser heating setup for diamond anvil cells for in situ synchrotron and in house high and ultra-high pressure studies},
  journal = {Rev. Sci. Instrum.},
  volume  = {90},
  number  = {10},
  pages   = {104501},
  year    = {2019},
  doi     = {10.1063/1.5117786}
}

@article{Garbarino2024,
  author  = {G. Garbarino and M. E. Hanfland and S. Gallego-Parra and A. D. Rosa and M. Mezouar and D. Duran and J. Jacobs},
  title   = {Extreme conditions X-ray diffraction and imaging beamline {ID15B} on the {ESRF} extremely brilliant source},
  journal = {High Press. Res.},
  volume  = {44},
  number  = {3},
  pages   = {199--216},
  year    = {2024},
  doi     = {10.1080/08957959.2024.2379294}
}

@article{Prescher2015,
  author  = {C. Prescher and V. B. Prakapenka},
  title   = {{DIOPTAS}: A program for reduction of two-dimensional X-ray diffraction data and data exploration},
  journal = {High Press. Res.},
  volume  = {35},
  number  = {3},
  pages   = {223--230},
  year    = {2015},
  doi     = {10.1080/08957959.2015.1059835}
}

@article{Petricek2023,
  author    = {V. Petricek and L. Palatinus and J. Plasil and M. Dusek},
  title     = {Jana2020 -- a new version of the crystallographic computing system Jana},
  journal   = {Zeitschrift f{\"u}r Kristallographie - Crystalline Materials},
  volume    = {238},
  number    = {7-8},
  pages     = {271--282},
  year      = {2023},
  doi       = {10.1515/zkri-2023-0005}
}

@article{Petricek2014,
  author    = {V. Petricek and M. Dusek and L. Palatinus},
  title     = {Crystallographic Computing System {JANA}2006: General features},
  journal   = {Zeitschrift f{\"u}r Kristallographie - Crystalline Materials},
  volume    = {229},
  number    = {5},
  pages     = {345--352},
  year      = {2014},
  doi       = {10.1515/zkri-2014-1737}
}

@article{Giefers2007,
  author  = {H. Giefers and F. Porsch},
  title   = {Shear induced phase transition in {PbO} under high pressure},
  journal = {Physica B},
  volume  = {400},
  number  = {1--2},
  pages   = {53--58},
  year    = {2007},
  doi     = {10.1016/j.physb.2007.06.033}
}

@article{Adams1992,
  author  = {D. M. Adams and A. G. Christy and J. Haines and S. M. Clark},
  title   = {Second-order phase transition in {PbO} and {SnO} at high pressure: Implications for the litharge--massicot phase transformation},
  journal = {Phys. Rev. B},
  volume  = {46},
  number  = {18},
  pages   = {11358--11367},
  year    = {1992},
  doi     = {10.1103/PhysRevB.46.11358}
}

@article{Sorrell1970thermal,
  author  = {C. A. Sorrell},
  title   = {Thermal expansion of orthorhombic {PbO}},
  journal = {J. Am. Ceram. Soc.},
  volume  = {53},
  pages   = {552--554},
  year    = {1970},
  doi     = {10.1111/j.1151-2916.1970.tb15964.x}
}

@article{Hill1985,
  author  = {R. J. Hill},
  title   = {Refinement of the structure of orthorhombic {PbO} (massicot) by Rietveld analysis of neutron powder diffraction data},
  journal = {Acta Crystallogr. C},
  volume  = {41},
  pages   = {1281--1284},
  year    = {1985},
  doi     = {10.1107/S0108270185007571}
}

@article{Cochran1960,
  author  = {W. Cochran},
  title   = {Crystal stability and the theory of ferroelectricity},
  journal = {Phys. Rev. Lett.},
  volume  = {3},
  number  = {9},
  pages   = {412--414},
  year    = {1959},
  doi     = {10.1103/PhysRevLett.3.412}
}

@article{Posternak1994,
  author  = {M. Posternak and R. Resta and A. Baldereschi},
  title   = {Role of covalent bonding in the polarization of perovskite oxides: The case of {K}{Nb}{O}$_3$},
  journal = {Phys. Rev. B},
  volume  = {50},
  number  = {12},
  pages   = {8911--8914},
  year    = {1994},
  doi     = {10.1103/PhysRevB.50.8911}
}

@article{Kornev2005B,
  author  = {I. A. Kornev and L. Bellaiche and P. Bouvier and P.-E. Janolin and B. Dkhil and J. Kreisel},
  title   = {Ferroelectricity of perovskites under pressure},
  journal = {Phys. Rev. Lett.},
  volume  = {95},
  number  = {19},
  pages   = {196804},
  year    = {2005},
  doi     = {10.1103/PhysRevLett.95.196804}
}

@article{Mazin1998,
  author  = {Igor I. Mazin and Yingwei Fei and Robert Downs and Ronald E. Cohen},
  title   = {Possible polytypism in {FeO} at high pressures},
  journal = {American Mineralogist},
  volume  = {83},
  number  = {5-6},
  pages   = {451--457},
  year    = {1998},
  doi     = {10.2138/am-1998-5-605}
}

@article{Lee2022,
  author  = {J. G. Lee and C. J. Pickard and B. Cheng},
  title   = {High-pressure phase behaviors of titanium dioxide revealed by a $\Delta$-learning potential},
  journal = {J. Chem. Phys.},
  volume  = {156},
  number  = {7},
  pages   = {074106},
  year    = {2022},
  doi     = {10.1063/5.0079844}
}

@article{Dubrovinsky2001,
  author  = {L. S. Dubrovinsky and N. A. Dubrovinskaia and V. Swamy and J. Muscat and N. M. Harrison and R. Ahuja and B. Holm and B. Johansson},
  title   = {The hardest known oxide},
  journal = {Nature},
  volume  = {410},
  pages   = {653--654},
  year    = {2001},
  doi     = {10.1038/35070650}
}

@article{AlKhatatbeh2009,
  author  = {Y. Al-Khatatbeh and K. K. M. Lee and B. Kiefer},
  title   = {High-pressure behavior of {TiO}$_2$ as determined by experiment and theory},
  journal = {Phys. Rev. B},
  volume  = {79},
  pages   = {134114},
  year    = {2009},
  doi     = {10.1103/PhysRevB.79.134114}
}

@article{Swamy2006,
  author  = {V. Swamy and A. Kuznetsov and L. S. Dubrovinsky and P. F. McMillan and V. B. Prakapenka and G. Shen and B. C. Muddle},
  title   = {Size-dependent pressure-induced amorphization in nanoscale {TiO}$_2$},
  journal = {Phys. Rev. Lett.},
  volume  = {96},
  number  = {13},
  pages   = {135702},
  year    = {2006},
  doi     = {10.1103/PhysRevLett.96.135702}
}

@article{Wang2001,
  author  = {Z. Wang and S. K. Saxena},
  title   = {Raman spectroscopic study on pressure-induced amorphization in nanocrystalline anatase ({TiO}$_2$)},
  journal = {Solid State Commun.},
  volume  = {118},
  number  = {2},
  pages   = {75--78},
  year    = {2001},
  doi     = {10.1016/S0038-1098(01)00046-1}
}

@article{Li2010,
  author  = {Q. Li and B. Liu and L. Wang and D. Li and R. Liu and B. Zou and T. Cui and G. Zou and Y. Meng and H. Mao and Z. Liu and J. Liu and J. Li},
  title   = {Pressure-induced amorphization and polyamorphism in one-dimensional single-crystal {TiO}$_2$ nanomaterials},
  journal = {J. Phys. Chem. Lett.},
  volume  = {1},
  number  = {1},
  pages   = {309--314},
  year    = {2010},
  doi     = {10.1021/jz9001828}
}

@article{akahama2006,
  author  = {Y. Akahama and H. Kawamura},
  title   = {Pressure calibration of diamond anvil Raman gauge to 310 {GP}a},
  journal = {J. Appl. Phys.},
  volume  = {100},
  number  = {4},
  pages   = {043516},
  year    = {2006},
  doi     = {10.1063/1.2335683}
}

@article{Mao1986,
  author  = {H. K. Mao and J. Xu and P. M. Bell},
  title   = {Calibration of the ruby pressure gauge to 800 kbar under quasi-hydrostatic conditions},
  journal = {J. Geophys. Res.},
  volume  = {91},
  number  = {B5},
  pages   = {4673--4676},
  year    = {1986},
  doi     = {10.1029/JB091iB05p04673}
}

@article{Teter1998,
  author    = {David M. Teter and Russell J. Hemley and Georg Kresse and J{\"u}rgen Hafner},
  title     = {High-pressure polymorphism in silica},
  journal   = {Phys. Rev. Lett.},
  volume    = {80},
  pages     = {2145--2148},
  year      = {1998},
  doi       = {10.1103/PhysRevLett.80.2145}
}

@article{Wang2013PbO,
  author = {Wang, Yonggang and Lin, Xiaohuan and Zhang, Hao and Wen, Ting and Huang, Fuqiang and Li, Guobao and Wang, Yingxia and Liao, Fuhui and Lin, Jianhua},
  title = {Selected-control hydrothermal growths of $\alpha$- and $\beta$-PbO crystals and orientated pressure-induced phase transition},
  journal = {CrystEngComm},
  volume = {15},
  pages = {3513--3516},
  year = {2013},
  doi = {10.1039/C3CE26916A}
}

@article{Hill2025,
  author = {Hill, R. J. and Cranswick, L. M. D.},
  title = {Crystallography of the litharge to massicot phase transformation from neutron powder diffraction data},
  journal = {Acta Crystallographica Section B: Structural Science, Crystal Engineering and Materials},
  volume = {81},
  pages = {146--160},
  year = {2025},
  doi = {10.1107/S205252062401254X}
}

@article{Racioppi2025,
  author       = {Stefano Racioppi and Alberto Otero-de-la-Roza and Samad Hajinazar and Eva Zurek},
  title        = {Powder {X}-ray diffraction assisted evolutionary algorithm for crystal structure prediction},
  journal      = {Digital Discovery},
  year         = {2025},
  volume       = {4},
  pages        = {73--83},
  doi          = {10.1039/D4DD00210K},
}

@article{Hajinazar2024,
  author       = {Samad Hajinazar and Eva Zurek},
  title        = {{X}tal{O}pt version 13: Multi-objective evolutionary search for novel functional materials},
  journal      = {Computer Physics Communications},
  year         = {2024},
  volume       = {304},
  pages        = {109306},
  issn         = {0010-4655},
  doi          = {10.1016/j.cpc.2024.109306},
}

@article{Lonie2011,
  author    = {Lonie, D. C. and Zurek, E.},
  title     = {Xtal{O}pt: An open-source evolutionary algorithm for crystal structure prediction},
  journal   = {Comput. Phys. Commun.},
  volume    = {182},
  pages     = {372},
  year      = {2011},
  doi       = {10.1016/j.cpc.2010.07.048}
}

@article{Avery2017,
  author    = {Avery, P. and Zurek, E.},
  title     = {Rand{S}pg: An open-source program for generating atomistic crystal structures with specific space groups},
  journal   = {Comput. Phys. Commun.},
  volume    = {213},
  pages     = {208},
  year      = {2017},
  doi       = {10.1016/j.cpc.2016.12.020}
}

@article{Lonie2012,
  author    = {Lonie, D. C. and Zurek, E.},
  title     = {Identifying duplicate crystal structures: {XtalC}omp, an open-source solution},
  journal   = {Comput. Phys. Commun.},
  volume    = {183},
  pages     = {690},
  year      = {2012},
  doi       = {10.1016/j.cpc.2011.11.008}
}

@article{MejiaRodriguez2020,
  author    = {Mej{\'i}a-Rodr{\'i}guez, D. and Trickey, S. B.},
  title     = {Meta-{GGA} performance in solids at almost {GGA} cost},
  journal   = {Phys. Rev. B},
  volume    = {102},
  pages     = {121109(R)},
  year      = {2020},
  doi       = {10.1103/PhysRevB.102.121109}
}

@article{Qi2023,
  author    = {Qi, W. and Xie, C. and Hushur, A. and Kojima, S.},
  title     = {Pressure-induced successive phase transitions and Fano resonance engineering in lead-free piezoceramics {KNbO}$_3$},
  journal   = {Appl. Phys. Lett.},
  volume    = {122},
  pages     = {232901},
  year      = {2023},
  doi       = {10.1063/5.0143105}
}

@article{Hemley1992_SilicatePerovskite,
  author    = {R. J. Hemley and R. E. Cohen},
  title     = {Silicate Perovskite},
  journal   = {Annu. Rev. Earth Planet. Sci.},
  year      = {1992},
  volume    = {20},
  pages     = {553--600},
}

@article{Hirose2006,
  author  = {K. Hirose},
  title   = {Postperovskite phase transition and its geophysical implications},
  journal = {Rev. Geophys.},
  volume  = {44},
  pages   = {RG3001},
  year    = {2006},
  doi     = {10.1029/2005RG000186}
}

@article{Samara1996,
 author  = {G.A. Samara},
 title   = {Pressure-Induced Crossover from Long- to Short-Range Order in Compositionally Disordered Soft Mode Ferroelectrics},
 journal = {Phys. Rev. Lett.},
 volume  = {77},
 pages   = {314-317},
 year    = {1996},
 doi     = {https://doi.org/10.1103/PhysRevLett.77.314}
}

@article{Young2006IrN2OsN2,
  title     = {Synthesis of Novel Transition Metal Nitrides {IrN}$_2$ and {OsN}$_2$},
  author    = {Young, Andrea F. and Sanloup, Chrystele and Gregoryanz, Eugene and
               Scandolo, Sandro and Hemley, Russell J. and Mao, Ho-kwang},
  journal   = {Phys. Rev. Lett.},
  volume    = {96},
  pages     = {155501},
  year      = {2006},
  doi       = {10.1103/PhysRevLett.96.155501}
}

@article{Vadapoo2017Oxynitride,
  title     = {Synthesis of a Polar Ordered Oxynitride Perovskite},
  author    = {Vadapoo, Rajasekarakumar and Ahart, Muhtar and Somayazulu, Maddury and
               Holtgrewe, Nicholas and Meng, Yue and Konopkova, Zuzana and
               Hemley, Russell J. and Cohen, R. E.},
  journal   = {Phys. Rev. B},
  volume    = {95},
  pages     = {214120},
  year      = {2017},
  doi       = {10.1103/PhysRevB.95.214120}
}

@article{Somayazulu2019LaH10,
  title     = {Evidence for Superconductivity above 260~{K} in Lanthanum Superhydride at Megabar Pressures},
  author    = {Somayazulu, Maddury and Ahart, Muhtar and Mishra, Ajay K. and
               Geballe, Zachary M. and Baldini, Maria and Meng, Yue and
               Struzhkin, Viktor V. and Hemley, Russell J.},
  journal   = {Phys. Rev. Lett.},
  volume    = {122},
  pages     = {027001},
  year      = {2019},
  doi       = {10.1103/PhysRevLett.122.027001}
}

@article{Birch1952Elasticity,
  author  = {Birch, Francis},
  title   = {Elasticity and Constitution of the {Earth}'s Interior},
  journal = {J. Geophys. Res.},
  volume  = {57},
  number  = {2},
  pages   = {227--286},
  year    = {1952},
  doi     = {10.1029/JZ057I002P00227}
}

@article{Burns1970PRL,
  author  = {Burns, G. and Scott, B. A. and Taylor, E. F.},
  title   = {Raman Scattering in Ferroelectric Perovskites},
  journal = {Phys. Rev. Lett.},
  volume  = {25},
  pages   = {167--170},
  year    = {1970},
  doi     = {10.1103/PhysRevLett.25.167}
}

@article{Park1997JAP,
  author  = {Park, S.-E. and Shrout, T. R.},
  title   = {Ultrahigh Strain and Piezoelectric Behavior in Relaxor Based Ferroelectric Single Crystals},
  journal = {J. Appl. Phys.},
  volume  = {82},
  pages   = {1804--1811},
  year    = {1997},
  doi     = {10.1063/1.365983}
}

\BeginSupplement

\setcounter{figure}{0}\renewcommand{\thefigure}{S\arabic{figure}}
\setcounter{table}{0}\renewcommand{\thetable}{S\arabic{table}}
\setcounter{equation}{0}\renewcommand{\theequation}{S\arabic{equation}}

\twocolumngrid

Additional technical details of the experimental and computational methods are provided here. Synchrotron X-ray diffraction was performed at ESRF (ID15B, $\lambda=0.4100$~\AA) and APS (16-ID-B, $\lambda=0.4133$~\AA). Polycrystalline PbTiO$_3$ (\(\geq\)99\%, $<$5~\textmu m, Sigma-Aldrich) was loaded into diamond anvil cells with 100~\textmu m culets and 50~\textmu m gasket holes. Selected high-temperature measurements employed one-sided laser heating. Diffraction images were integrated in DIOPTAS~\cite{Prescher2015}, and Le Bail refinements carried out in JANA~\cite{Petricek2014,Petricek2023}. Pressures were calibrated by ruby fluorescence and the diamond Raman edge~\cite{Mao1986,akahama2006}.

Density Functional Theory (DFT) calculations were performed using VASP, version 6.4.2~\cite{Kresse1994}. The r$^2$SCAN~\cite{Furness2020} exchange-correlation functional was employed for the geometry optimizations and calculations of the electronic properties. The projector augmented wave (PAW) method~\cite{Blochl1994}, with a cutoff of 700~eV, was used. The Ti $3s^23p^64s^23d^2$ (PAW\_PBE Ti\_sv), Pb $5d^{10}6s^26p^2$ (PAW\_PBE Pb\_d) and O $2s^22p^4$ (PAW\_PBE O\_s) states were treated explicitly. 

The \textbf{k}-point meshes were generated using the $\Gamma$-centered Monkhorst--Pack scheme~\cite{Monkhorst1976}, with the number of divisions along each reciprocal lattice vector chosen such that the product of this number with the real lattice constant was $\geq 60$~\AA. For geometry optimizations, the energy convergence criterion was $10^{-6}$~eV and the norms of all forces were smaller than $10^{-3}$~eV/\AA. Gaussian smearing with a width of 0.1~eV was adopted during geometry optimizations, and the tetrahedron smearing method~\cite{Blochl1994_tetra} was used to calculate the electronic properties of optimized geometries. The non-spherical contributions related to the gradient of the density in the PAW spheres were included.

Phonons in the harmonic approximation were determined for the PbO phases with the Phonopy package~\cite{Togo2008}, using supercells based on the conventional unit cells of the r$^2$SCAN-optimized structures. A finite displacement of 0.003~\AA{} was applied, and the energy convergence criterion was set to $10^{-8}$~eV.

Crystal structure prediction (CSP) searches were carried out using the open-source evolutionary algorithm \textit{XtalOpt} (version~13.0)~\cite{Lonie2011,Hajinazar2024}. The multi-objective fitness measure was employed in combination with the powder X-ray diffraction (PXRD)–assisted module~\cite{Racioppi2025} to evaluate structural agreement with the experimental PXRD data. The absence of matches with PbTiO$_3$ indicated possible thermal decomposition, which was subsequently confirmed through additional searches on PbO.

The initial generation consisted of random symmetric structures that were created by the RandSpg algorithm~\cite{Avery2017}. The number of initial structures was equal to 200. The number of formula units (FUs) was set equal to 1, 2, 3, 4, 6, and 8. A weight parameter of 0.7 was used for the similarity index in the CSP searches. Duplicate structures were identified and removed from the breeding pool using the XtalComp algorithm ~\cite{Lonie2012}. The total number of generated structures per run was equal to 1000. The structure search followed a multi-step strategy, with three subsequent optimizations with increased levels of accuracy. Geometry optimizations in the CSP search were performed using DFT at 80 GPa, in combination with the meta-GGA exchange-correlation functional r$^2$SCAN-L~\cite{MejiaRodriguez2020}. Also for the CSP, the k-point meshes were generated using the $\Gamma$-centered Monkhorst-Pack scheme, and the number of divisions along each reciprocal lattice vector was selected so that the product of this number with the real lattice constant was greater than or equal to a given cutoff. The values of 30, 40, and 50~\AA{} were used for the three subsequent optimization steps in the crystal structure search. The accuracy of the energy convergence was set to increase from $10^{-3}$~eV to $10^{-5}$~eV eV for the optimizations. 

\begin{figure}[htbp]
    \centering
    \includegraphics[width=0.8\linewidth]{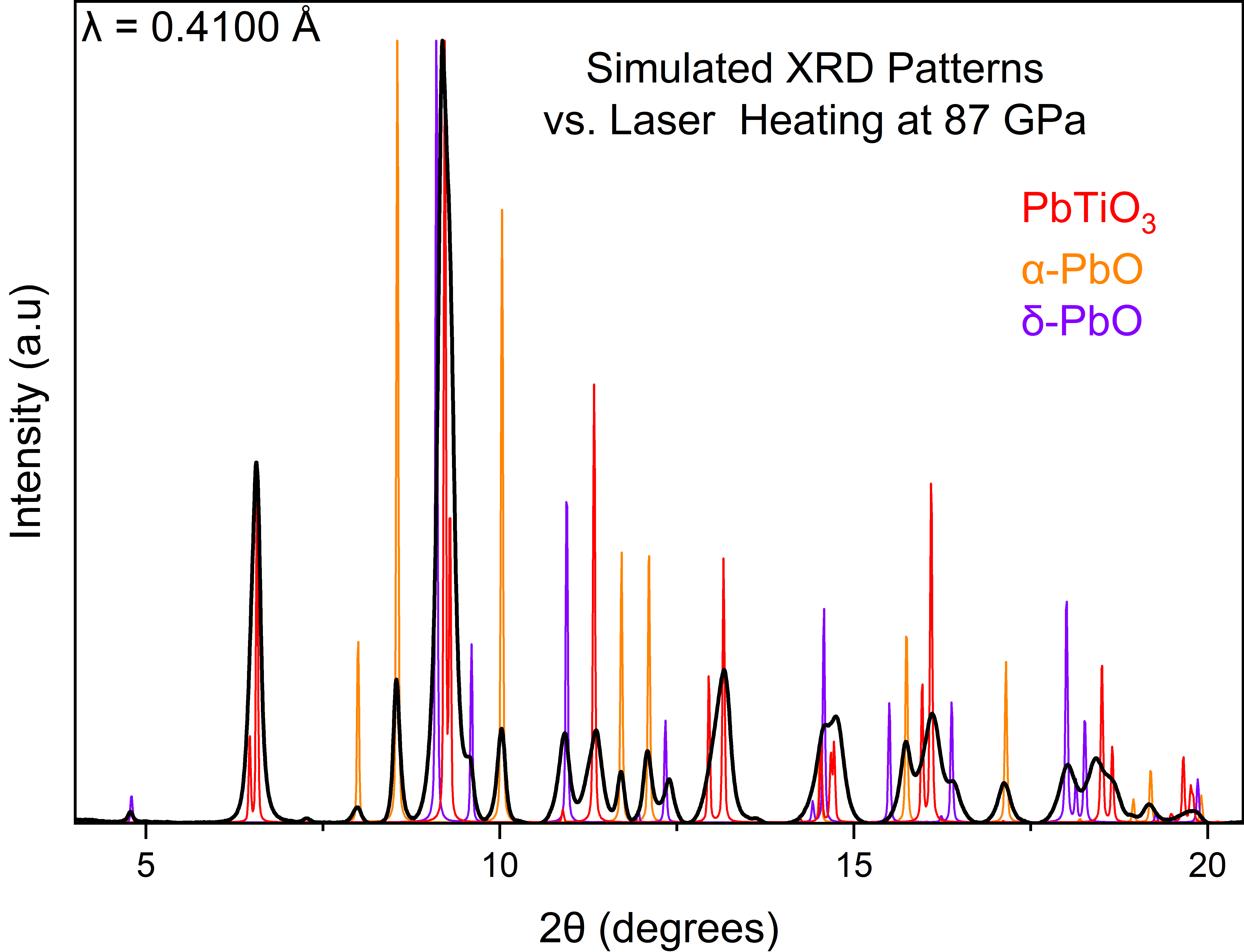}
    \caption{Simulated XRD patterns at 87~GPa over the laser-heated sample of PbTiO$_3$.}
    \label{fig:S5}
\end{figure}

\begin{table}[htbp]
\caption{Refined lattice parameters and unit-cell volumes of PbTiO$_3$ and PbO phases ($\alpha$-PbO and $\delta$-PbO) at 87~GPa (laser-heated sample).}
\label{tab:lattice_hp}
\begin{tabular}{lccc}
\hline\hline
Phase & $a$ (\AA) & $c$ (\AA) & Volume (\AA$^3$) \\
\hline
PbTiO$_3$ (\textit{I}4/\textit{mcm}) & 5.06 & 7.27 & 186.0 \\
$\alpha$-PbO (\textit{P}4/\textit{nmm}) & 3.89 & 2.94 & 44.6 \\
$\delta$-PbO (\textit{P}4/\textit{nmm}) & 3.04 & 4.90 & 45.2 \\
\hline\hline
\end{tabular}
\end{table}

\begin{table}[htbp]
\caption{Refined lattice parameters and unit-cell volumes of PbTiO$_3$ and $\beta$-PbO near ambient pressure (post-decompression).}
\label{tab:lattice_decomp}
\begin{tabular}{lcccc}
\hline\hline
Phase & $a$ (\AA) & $b$ (\AA) & $c$ (\AA) & Volume (\AA$^3$) \\
\hline
PbTiO$_3$ (\textit{P}4\textit{mm}) & 3.91 & 3.91 & 4.15 & 63.46 \\
$\beta$-PbO (\textit{Pbcm)} & 5.80 & 5.46 & 4.66 & 147.72 \\
\hline\hline
\end{tabular}
\end{table}
\FloatBarrier

\begin{table}[htbp]
\caption{Refined lattice parameters and unit-cell volumes of PbTiO$_3$ and $\beta$-PbO near ambient pressure (post-decompression).}
\label{tab:lattice_decomp}
\begin{tabular}{lcccc}
\hline\hline
Phase & $a$ (\AA) & $b$ (\AA) & $c$ (\AA) & Volume (\AA$^3$) \\
\hline
PbTiO$_3$ (\textit{P}4\textit{mm}) & 3.91 & 3.91 & 4.15 & 63.5 \\
$\beta$-PbO (\textit{Pbcm)} & 5.80 & 5.46 & 4.66 & 147.7 \\
\hline\hline
\end{tabular}
\end{table}
\FloatBarrier

\begin{figure}[httb]
    \includegraphics[width=0.91\linewidth]{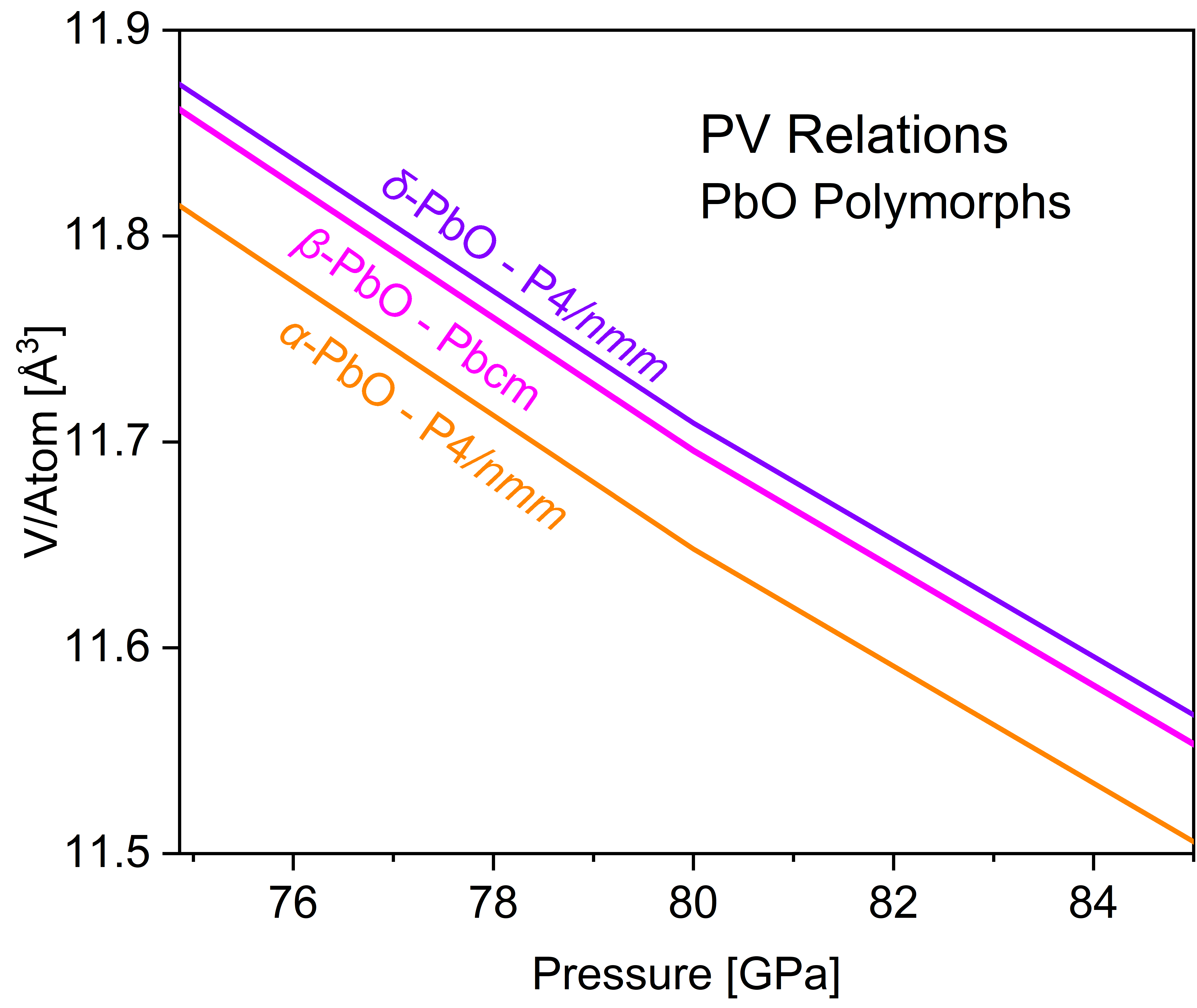}
    \caption{\textit{P–V} relations of PbO polymorphs based on DFT calculations showing their similar compressibility and close densities}
    \label{fig:PbO_PV}
\end{figure}

\FloatBarrier
\begin{figure}[httb]
    \includegraphics[width=0.91\linewidth]{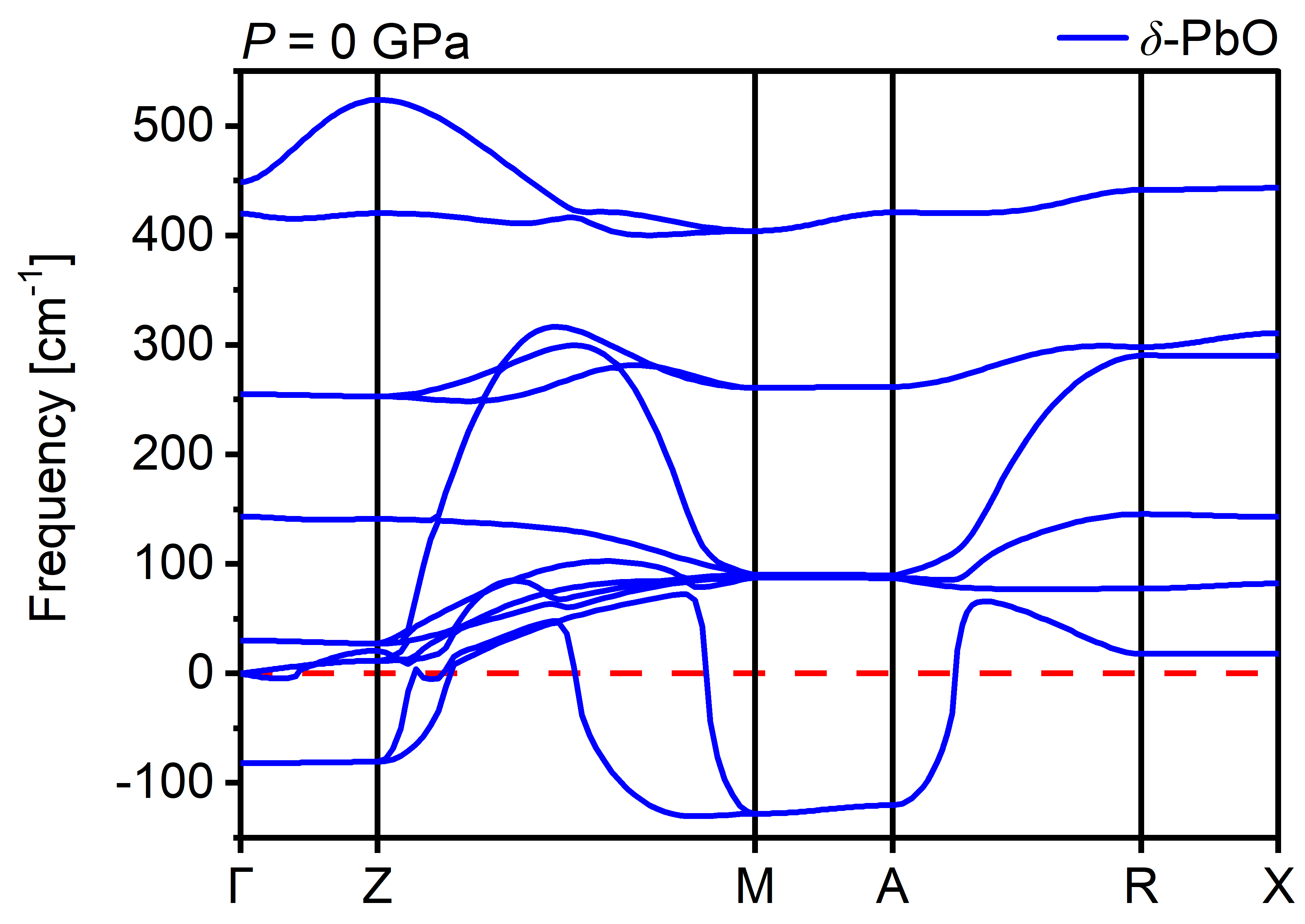}
    \caption{Phonon dispersion of $\delta$-PbO at zero pressure.}
    \label{fig:S1}
\end{figure}

\FloatBarrier
\begin{figure}[httb]
    \includegraphics[width=0.91\linewidth]{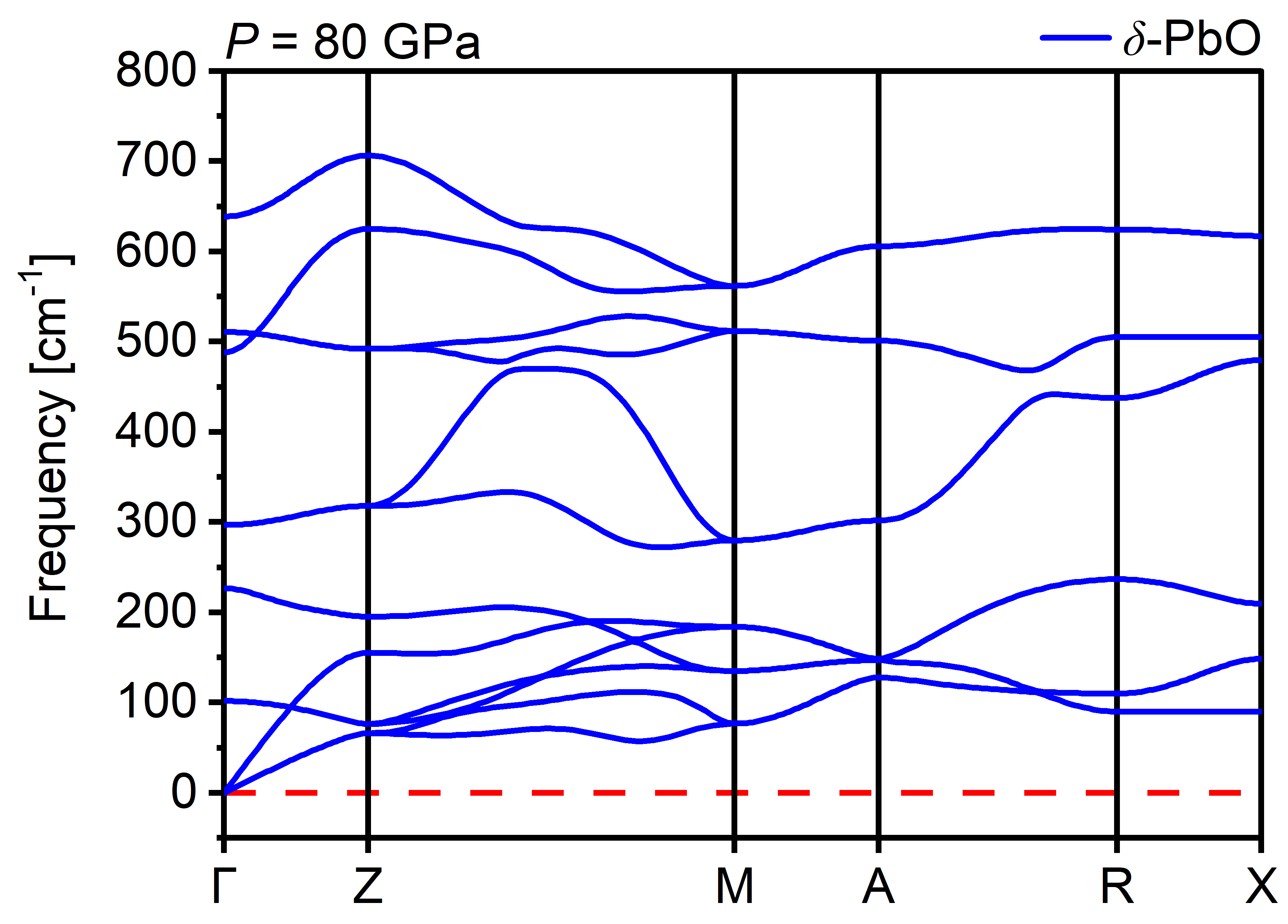}
    \caption{Phonon dispersion of $\delta$-PbO at 80~GPa.}
    \label{fig:S1b}
\end{figure}
\FloatBarrier
\begin{figure}[httb]
    \includegraphics[width=0.91\linewidth]{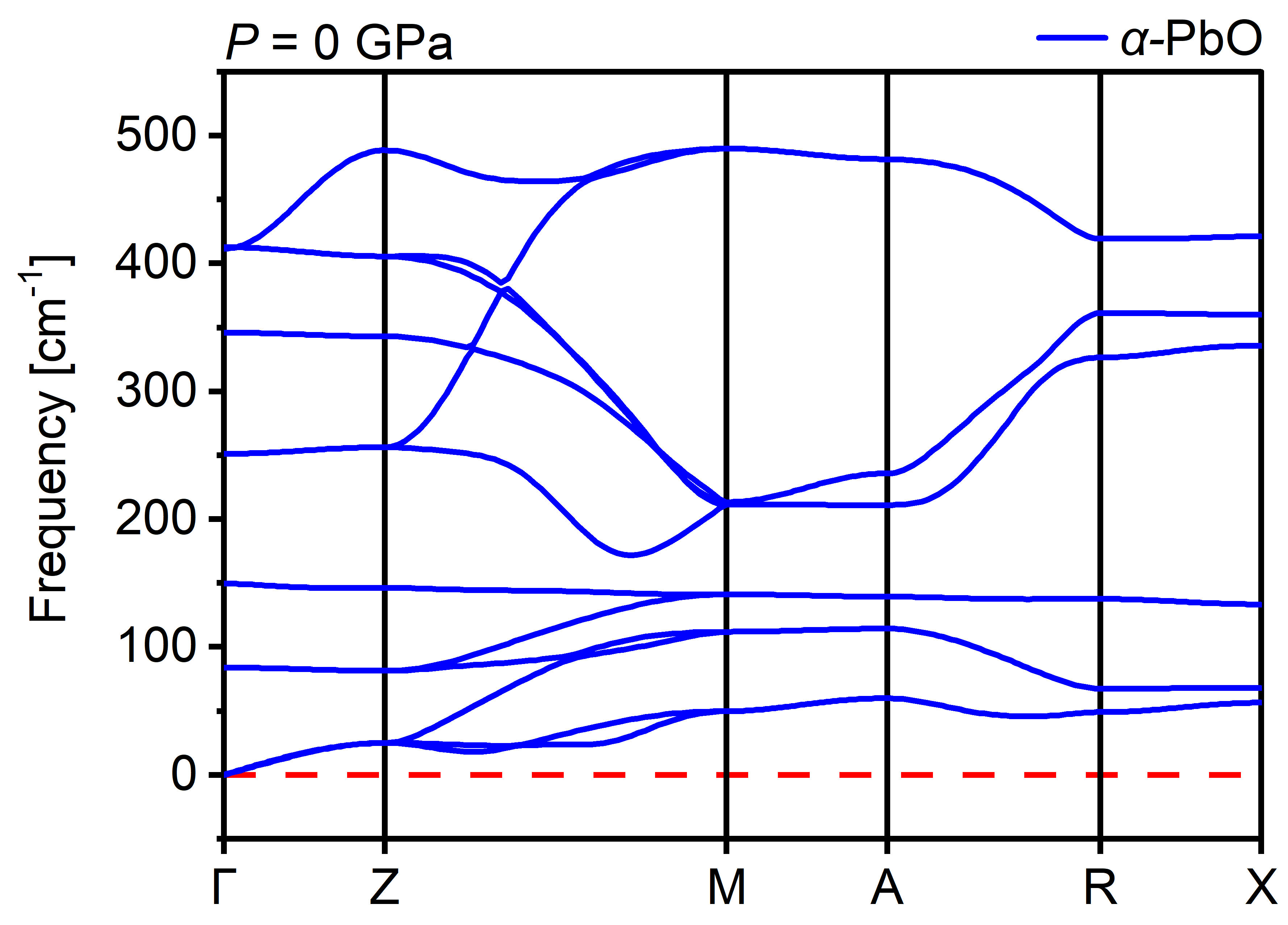}
    \caption{Phonon dispersion of $\alpha$-PbO at zero pressure.}
    \label{fig:S2}
\end{figure}
\FloatBarrier
\begin{figure}[httb]
    \includegraphics[width=0.91\linewidth]{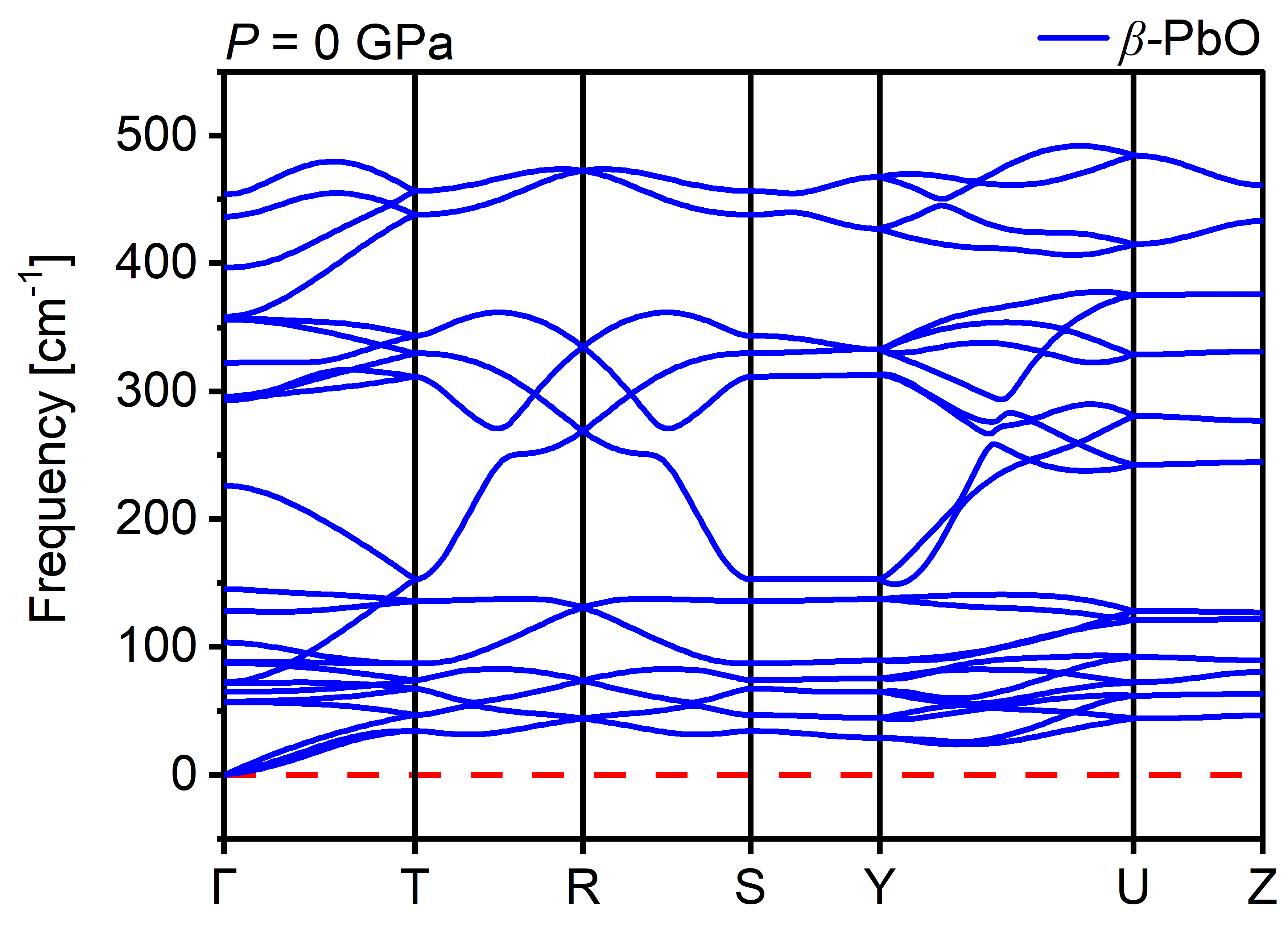}
    \caption{Phonon dispersion of $\beta$-PbO at zero pressure.}
    \label{fig:S3}
\end{figure}
\FloatBarrier

\begin{figure}[httb]    
    \includegraphics[width=\linewidth]{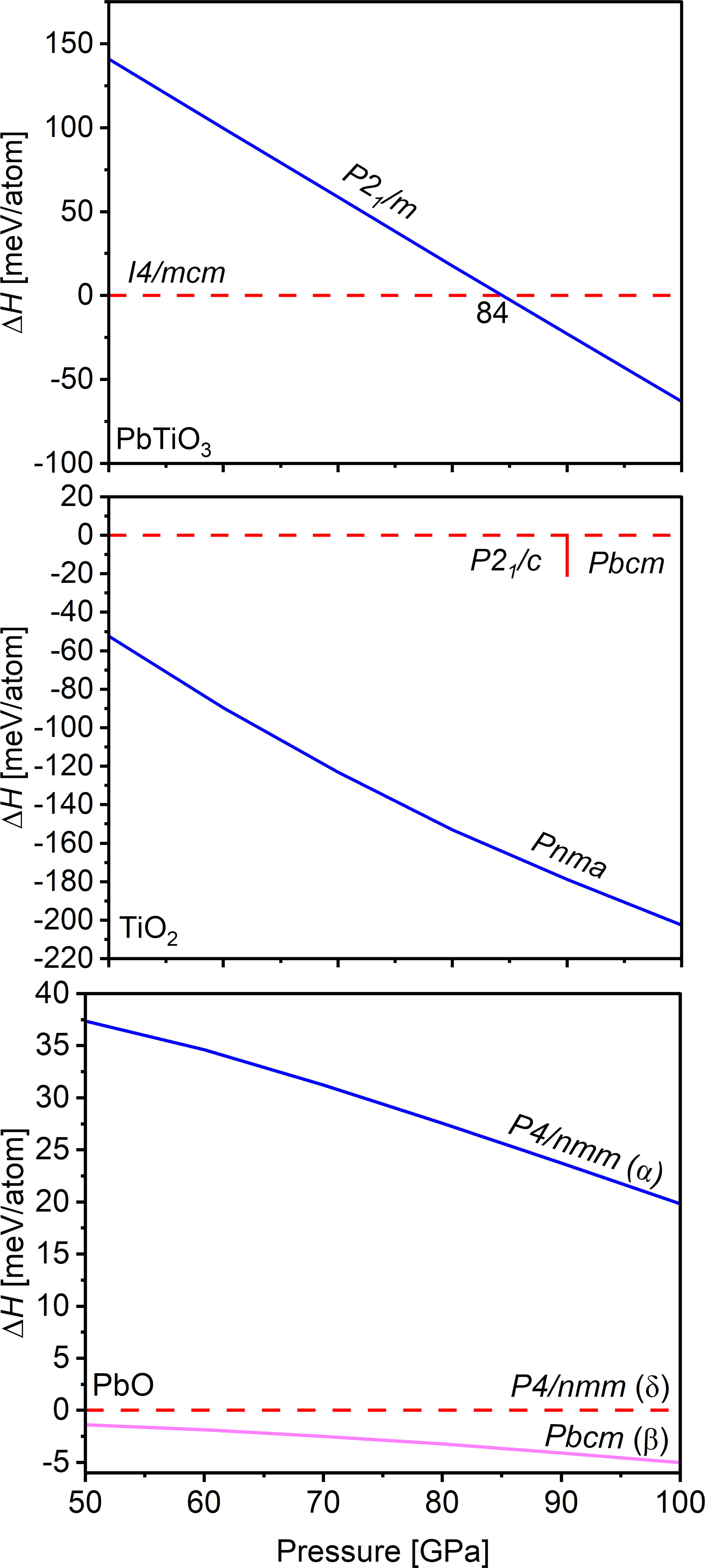}
    \caption{Calculated enthalpy difference ($H = E + PV$) as a function of pressure for PbTiO$_3$, TiO$_2$, and PbO. The red vertical line in the middle inset highlights the $P$2$_1$$/c$ to $Pbcm$ phase transition in TiO$_2$.}
    \label{fig:S4}
\end{figure}

\FloatBarrier

\end{document}